\newcommand{\beq}{\begin{equation}}
\newcommand{\eeq}{\end{equation}}

\newcommand{\bear}{\begin{eqnarray}}
\newcommand{\eear}{\end{eqnarray}}

\newcommand{\tn}{\textnormal}

\documentclass{JINST}

\usepackage{amssymb}

\title{Performances of multi-gap timing RPCs for relativistic ions in the range Z=1-6}

\author{P. Cabanelas$^a$, M. Morales$^a$, J. A. Garzon$^a$, A. Gil$^b$, 
D. Gonzalez-Diaz$^c$\thanks{Corresponding author}, A. Blanco$^d$, D. Belver$^a$, 
E. Casarejos$^d$, P. Fonte$^d$$^e$, W. Koenig$^c$, L. Lopes$^d$, M. Palka$^c$, 
J. Pietraszko$^c$, M. Traxler$^c$ and M. Weber$^f$\\
\llap{$^a$} Universidade de Santiago de Compostela (USC), Santiago de Compostela, Spain \\
\llap{$^b$} Instituto de F\' \i sica Corpuscular (CSIC-UV), Valencia, Spain \\
\llap{$^c$} Gesellschaft f\"ur Schwerionen Forschung (GSI), Darmstadt, Germany \\
\llap{$^d$} Laboratorio de Instrumenta\c{c}\~{a}o e F\' \i sica Experimental de Part\' \i culas (LIP), Coimbra, Portugal\ \
\llap{$^e$} Instituto Superior de Engenhar\' \i a de Coimbra (ISEC), Coimbra, Portugal \\
\llap{$^f$} Technische Universit\"at M\"unchen (TUM), Munich, Germany \\
E-mail: \email{D.Gonzalez-Diaz@gsi.de}}

\abstract{We present the performance of Multi-gap timing RPCs under irradiation by fully stripped relativistic ions 
($\gamma \beta$=2.7, $Z$=$1$-$6$). A time resolution of 80 ps at high efficiency
has been obtained by just using standard `off the shelf' 4-gap timing RPCs from the new HADES ToF wall. 
The resolution worsened to 100 ps for $\sim 1$ kHz/cm$^2$ proton flux and for $\sim 100$ Hz/cm$^2$ Carbon flux. 
The chambers were operated at a standard field of $E=100$ kV/cm and showed 
a high stability during the experiment, supporting the fact that RPCs are a convenient choice when 
accommodating a very broad range of ionizing particles is needed.

The data provides insight in the region of very highly ionizing particles (up to $\times 36$ mips)
and can be used to constrain the existing avalanche and Space-Charge models far from the 
usual `mip valley'. The implications of these results for the general case of detection based on
secondary processes (n, $\gamma$) resulting in highly ionizing particles 
with characteristic energy distributions will be discussed, together with the nature of the 
time-charge correlation curve.}

\keywords{multi-gap timing RPCs; time of flight; ion detection; highly ionizing particles}

\begin{document}

\section{Introduction}
\label{intro}

The development of large-scale sub-100 ps resolution RPCs has launched
a number of applications in Particle and Hadron Physics, notably
 HARP (PS) and ALICE (LHC), STAR (RHIC), FOPI, HADES (SIS18) and CBM (SIS100)
\cite{HARP, ALICE, STAR, FOPI, Paulo_HADES, CBM}. The rapid progress of
the field ever since its birth \cite{Pestov, Santonico} largely lies on the invention of the
multi-gap \cite{multigap}, the improvement in the gap definition quality 
that allowed to build an RPC at the 100 ps-$\sigma$ scale for the first time \cite{timing}
and the extension of the concept to long counters up to 1.6 m \cite{large}. Detectors of this
family are sometimes referred as Multi-gap timing RPCs\footnote{Also simply Multi-gap RPCs or timing RPCs.
Nevertheless, it must be mentioned that high time resolution is not a property of the Multi-gap
technology while high efficiency is also not a property of single gap timing RPCs.},
hereafter denoted simply by MtRPCs. All the so far existing
large walls based on MtRPCs emphasize the detector response for mips at moderate particle fluxes
(below 1 kHz/cm$^2$) that are typical environments either in high energy collider experiments or low
energy fixed-target ones.

There is, nonetheless, a broad range of newly born applications where MtRPCs need to work in
highly ionizing environments. Maybe the most remarkable one is the detection of annihilation $\gamma$'s
for Positron Emission Tomography (PET) \cite{PET} (based on the detection of the secondary electron (by Compton
or photo-electric effect)). More recently, in the framework of the R$^3$B collaboration, 
two main projects started
R\&D aimed at building large ToF walls for neutron and ion detection, respectively \cite{R3B,Kike}.
Both of them will have to deal with a yet unexplored range of highly ionizing particles.

Remarkable success has been achieved in describing the behavior of these chambers through first
principle avalanche simulations, revealing the main role of a very strong avalanche Space-Charge
\cite{Fonte_Q, Riegler_nice, Abb_simm}. Despite the presence of this complicated phenomena, a
handful of analytical expressions can be obtained if standard avalanche evolution is
assumed to happen up to the threshold level \cite{Abb_simm, Alessio1,Alessio2,Alessio3,Werner_summary}.
The developed formalisms allow to estimate the influence of the ionization loss, hereafter referred in
mips units ($\Delta{E}/\Delta{E_{mips}}$) as:
\bear
&&\bar{t}(\Delta{E}) = \frac{t_{rise}}{\ln{9}}\ln{\frac{\Delta{E}_{mips}}{\Delta{E}}}
+ t_{off}(n_{th}, t_{rise}) \label{to} \\ 
&&\sigma_{_T}(\Delta{E}) = K_1 \frac{t_{rise}}{\ln{9}} \sqrt{\frac{\Delta{E}_{mips}}{\Delta{E}}} 
\label{sigma_T} \\
&&\varepsilon(\Delta{E})\! =\! 1\!-\!\exp\!\left[{-n_o \left( 1-\frac{\eta}{\alpha} - 
\frac{\ln({1+\frac{(\alpha-\eta)}{E_w} n_{th}})}{\alpha g}\right) 
\frac{\Delta{E}}{\Delta{E}_{mips}}}\right] \label{eq_eff}
\eear
$\bar{t}$ refers to the average and $\sigma_{_T}$ to the rms of the time response
distribution (for the sake of simplicity it has been 
assumed to be Gaussian,
so that $\bar{t}$ can be thought of as the maximum of the time response distribution
and $\sigma_{_T}$ equals its rms).
$n_{th}$ is the electronics threshold in number of electrons, $n_o$ is the total number of primary ionizations
for mips, $g$ the gap size, $\alpha$ and $\eta$ the multiplication and attachment coefficients, $E_w$ 
the weighting field
 and $K_1$ is an adimensional constant of order unity that contains the effect of avalanche 
multiplication statistics. The signal rise-time $t_{rise}$ 
is related to the coefficients of the electron swarm as:
\beq
t_{rise}= \frac{\ln{9}}{(\alpha-\eta) v_e}
\eeq
in absence of Space-Charge (see \cite{Alessio3} for a discussion).
The average electron drift velocity is denoted by $v_e$. 
Typical values of $t_{rise}\simeq 200$-$250$ ps at
threshold level have been measured with a careful setup \cite{Paulo_IEEE, Diego_thesis},
for typical operating fields of $E=100$ kV/cm, and threshold levels of $q_{th}=n_{th} q_e \simeq 10$ fC. 
It must be noted that, within the models,
eq. \ref{to} is generally exact, being eq. \ref{sigma_T} full-filed in the limit
$\Delta{E}/{\Delta{E_{mips}}}\rightarrow\infty$ and, contrary, eq. \ref{eq_eff} for
$\Delta{E}/{\Delta{E_{mips}}}\rightarrow 1$ (otherwise the latter must be re-interpreted as a lower limit 
\cite{Werner_summary}).
Despite the development of analytical tools, no systematic attempt was done to clarify these
dependences with exception of the controverted data from \cite{HARP2} and \cite{HARP3}, 
that will be re-visited here, and the much too short survey of \cite{Alessio1} (only 2 points).

The detection of ions 
up to $A=200$ ($Z\simeq100$) over large surfaces ($\simeq 5$ m$^2$)
at relativistic kinetic energies $E_K=700$ GeV/A has recently brought attention in the
R$^3$B experiment, proposed within the new Facility for Anti-proton and Ion Research (FAIR)
at Darmstadt, Germany \cite{R3B}. In view of this potential new application,
we conducted systematic measurements at GSI-SIS18 for evaluating the detector response under
ions up to charge state $Z=6$. 
Complementary, measurements with a highly mono-energetic diffuse proton beam
($E_K=1.76$ GeV, $\sigma_{E_K}/E_K=4\%$) and cosmic rays were also performed. 
Spare cells from the HADES system were
used \cite{Alberto} and comparisons to previous
results will be referred when appropriate.

Importantly, paralell plate geometries (of the kind of PPACs) have been used for many years
for detecting ions at the 100 MeV energy scale or below with great success \cite{Stelzer, Breskin, Salabura}. 
Apart from extending the time resolution below 100 ps, the introduction of the RPC technology should
greatly improve the chamber stability and easy its construction, being intrinsically `spark-protected' 
at ambient pressure. As shown in this work, these features are already provided by typical MtRPC 
designs if ions have enough energy to penetrate in a relatively bulky detector ($\simeq 20\%X_o$ here)
and ion fluxes are below 1 kHz/cm$^2$.

The structure of the paper is as follows: the three different experimental setups are
explained in section \ref{intro}, the behavior as a function of HV and rate is
 presented in section \ref{results}, while section \ref{Z_dependence}
is devoted to the behavior as a function of particle type, a discussion follows in 
section \ref{Discussion} about the practical
use of such a counter in a highly ionizing environment and finally in 
section \ref{Conclusion} we summarize our conclusions.

\section{Experimental setup}
\label{intro}
\subsection{The RPC cells}

\begin{figure}[ht!!!]
\begin{center}
\includegraphics[width = \linewidth]{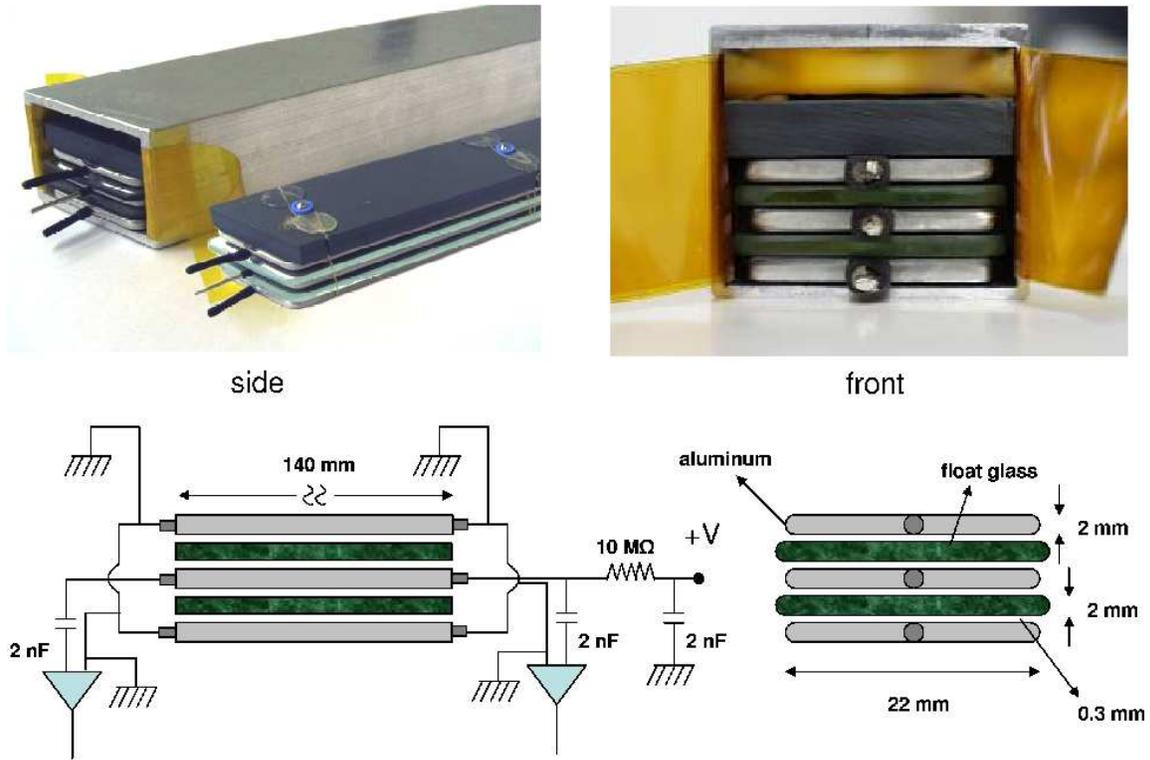}
\caption{\footnotesize Up: pictures showing a typical HADES cell, before and after being placed in its
corresponding Faraday cage. Down: side and front sketch of a HADES cell together with the electrical
scheme used for these measurements.}
\label{cell_zoom}
\end{center}
\end{figure}

Two RPC cells from the new HADES MtRPC wall 
 have been used and allocated inside their corresponding shielding profiles (acting as Faraday cages)
in a custom designed aluminum gas box.
As compared with the HADES system, the gas box was lacking of an optimized PCB for signal
feed-through of $Z_o = 50 \Omega$ (characteristic impedance) while the routing of the signals inside the
box was such that they could be read out from the same box side. Electrical isolation of signal
cables was ensured whenever possible in order to reduce cross-talk.
A picture of a typical cell is shown in Fig. \ref{cell_zoom}, consisting of
4 gaps of 0.28 mm thickness, with 2 mm thick aluminum and float glass plates. The
dimensions of the cells were $22 \times 140$ mm$^2$
(width $\times$ length), corresponding to the low polar angle region
of the HADES wall. Technical details can be found in \cite{Paulo_HADES}.

The read-out was based on the HADES FEE electronics that provides x40 amplification factor
in signal amplitude at 2 GHz bandwidth \cite{Dani_HADES}. The detector
was read out in single-ended mode, with a signal being taken from the central electrode (anode) after
filtering the HV level. No special care was taken for matching the detector impedance
($Z_d = 12 \pm 2 \Omega$) to the coaxial LEMO cables used for signal transmission ($Z_o=50 \Omega$),
 while the FEE built-in dead-time of the order of $50$-$70$ ns
provided stability to the electronic chain by avoiding re-triggers caused by reflections. 
 2 MBs (Mother-Boards) and 2x2 DBs (Daughter-Boards) \cite{Alex_HADES} fed by
a customized low-ripple distributed LV system \cite{Alex_LV} were used for reading-out
8 electronic channels ($2\times2$ RPC cells and $2\times2$ reference scintillators).
The RPC-FEE thresholds were set to $v_{th}=50$ mV 
(equivalently $v_{th}=50/40=1.25$ mV and $q_{th}\simeq 30$ fC at the pre-amplifier input) 
and were not changed during the measurements. The gas box was used as a 
central ground node being the FEE, HV and cathode strips connected to it by screws and/or conductive meshes.
With this grounding scheme the external noise levels were still modest and indeed for $v_{th}<20$ mV 
some channels were unstable.
The digitization and event building was done with the Trigger and Readout Board (TRB) developed 
in \cite{TRB} and based on the HPTDC ASIC chip. 

The oxygen content measured at the output of the gas box was below 200 ppms
 (measured with an Oxygen Transmitter O2X1 of GE Infrastructure Sensing) for a 
gas flow of 150 cc/min of an isobutane-free gas mixture based on C$_2$H$_2$F$_4$/SF$_6$ (90/10). 
Due to the high electro-negativity of the gas mixtures used for timing such ppm levels are not
expected to influence the detector behavior. 

\subsection{The trigger and reference system}
\label{trigger}
\begin{figure}[ht!!!]
\begin{center}
\includegraphics[width = 13 cm]{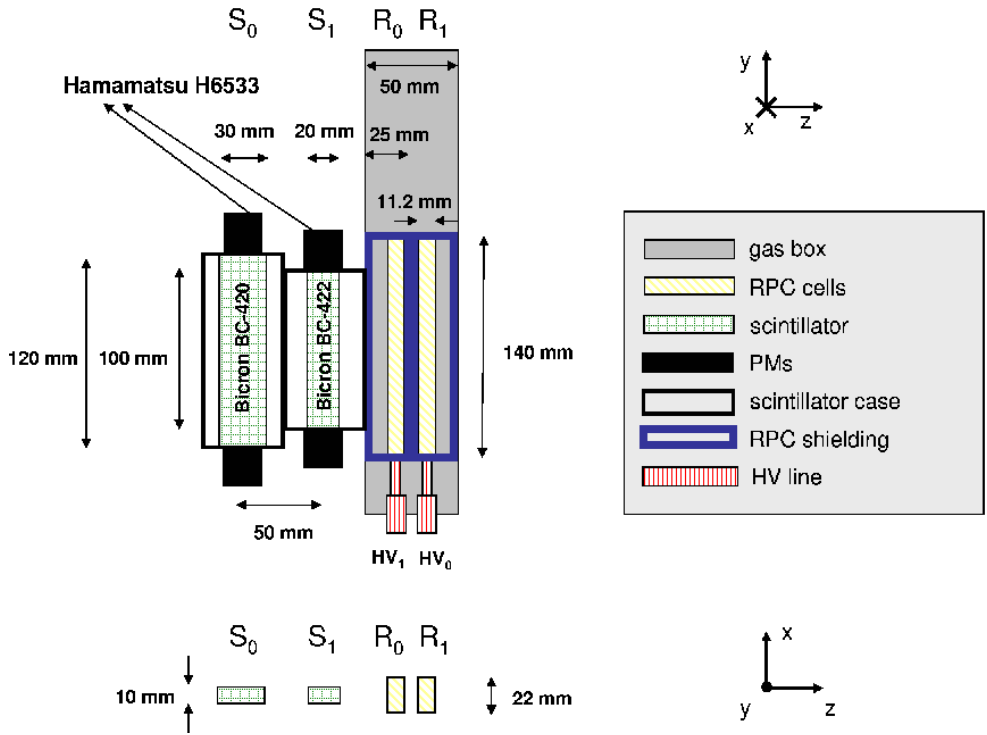}
\caption{\footnotesize Different views of the arrangement of the reference scintillators and RPC cells used
for the measurements.}
\label{setup_zoom}
\end{center}
\end{figure}

The arrangement of the RPC cells and reference scintillators is shown in Fig. \ref{setup_zoom}. Two fast
Bicron scintillators BC-420 and BC-422 read out in both ends by Hamamatsu H6533 photo-multipliers (PMs)
were used for providing trigger, reference time and track selection for efficiency studies.
The scintillator-PM assembly was placed inside PVC tubes of 50 mm diameter and
optical silicone was applied to the contact surface scintillator-PM while the scintillator was wrapped
in silver paper to improve light transmission. Such
a detector choice had previously yielded $\sigma_{_T} = 35$ ps per counter for mips \cite{large}.

For the scintillators readout the HADES FEE was also used with the
pre-amplifier being by-passed in virtue of the PM amplification. The PMs were 
operated at a nominal voltage $V=-2.2$ kV for which the average signal amplitude was just slightly higher 
than that of RPC signals after amplification. Variations
around the nominal voltage were introduced to account for slight differences
in the characteristic gain curves of the different PMs. The same threshold as in the RPC-FEE was used.

We have chosen a setup with the 4 counters (2 RPCs and 2 scintillators)
being placed vertically along their longest side. The scintillators, being prisms of dimensions
$10\times{20}\times{100}$ mm$^3$ (BC-422) and $10\times{30}\times{120}$ mm$^3$ (BC-420) were stacked over
their narrower faces ($10\times 100$ mm$^2$ and $10\times 120$ mm$^2$, respectively).
They could be aligned with an estimated precision of the order of $\pm 2$ mm.
A similar procedure was followed with the RPC cells.
Being mechanically identical and placed inside aluminum profiles,
both RPC cells can be easily stacked with a mutual alignment better than 1 mm. The reference system
was carefully centered with respect to the RPCs in order to provide a good reference for efficiency
estimates. Requesting an additional coincidence with one of the
(`reference' in such case) RPC is expected to compensate for any error on the aforementioned procedure
and was done when possible (measurements of section \ref{proton_beam}).

For all the measurements the trigger was provided by a coincidence of  the 4 PM output signals.
In order to do that the PM outputs were split, with one being sent to a Leading
Edge Discriminator (LED) while the other was fed directly
into the FEE after by-passing the amplifier, both discriminators having the same threshold.

Additionally to the scintillator reference system, two fast position sensitive mono-crystalline diamonds
were placed 14 m upstream,
roughly at the focal plane of the last
HADES beam-line quadrupole. The surface dimensions of the detectors
were $4.7\times4.7$ mm$^2$ and $3.5\times3.5$ mm$^2$ and their individual time resolutions in the range
$100$-$150$ ps \cite{Wolfgang}. Although not in the trigger because of technical reasons, a coincidence
probability of 10\% was observed off-line, allowing for an improved track selection and energy spread
determination. Those were used only in a reduced set of the measurements (section \ref{proton_beam}).

\subsection{Prompt charge determination}
\label{Q_cal}
\begin{figure}[ht!!!]
\begin{center}
\includegraphics[width = 6.7 cm]{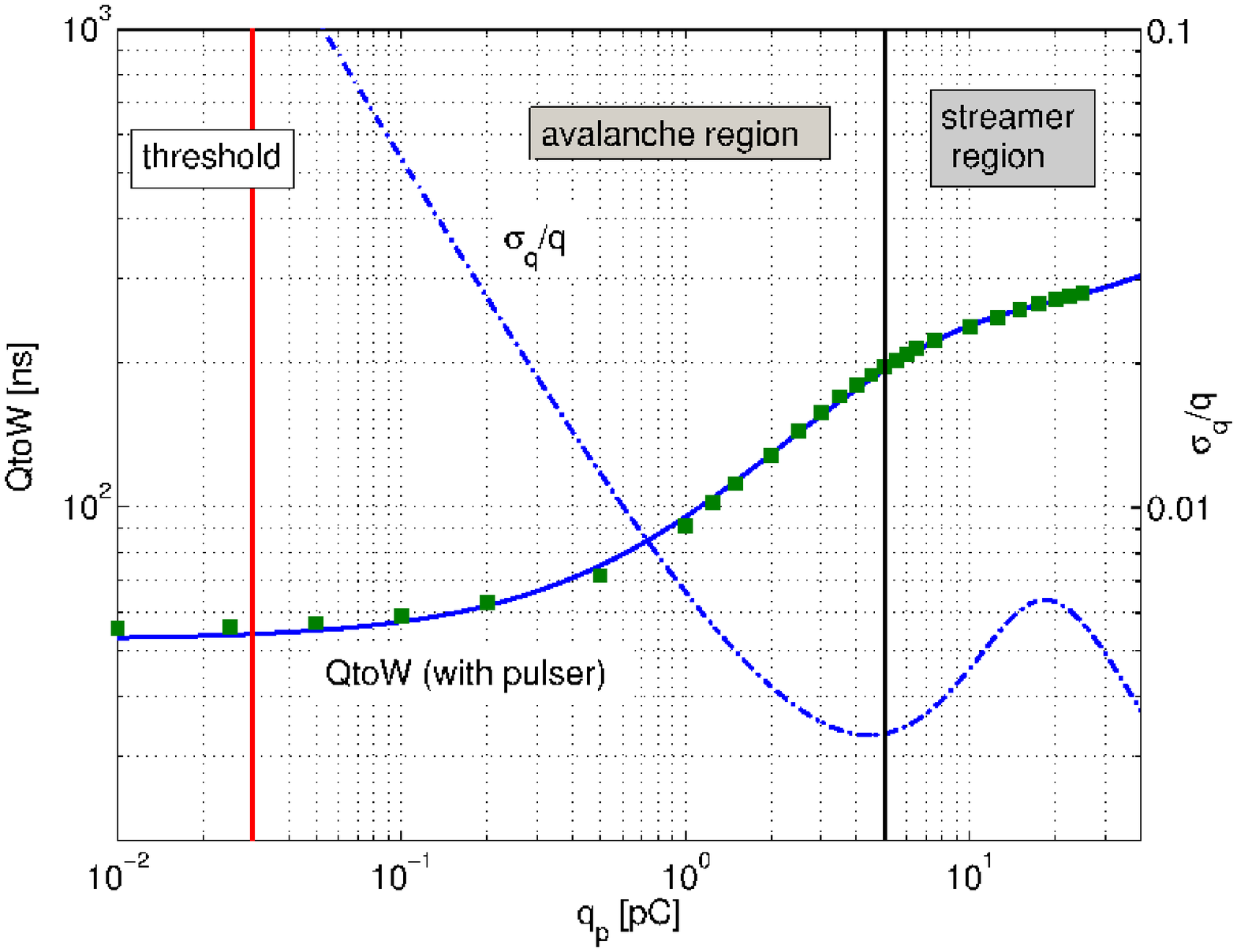}
\includegraphics[width = 6.9 cm, height=5.0 cm]{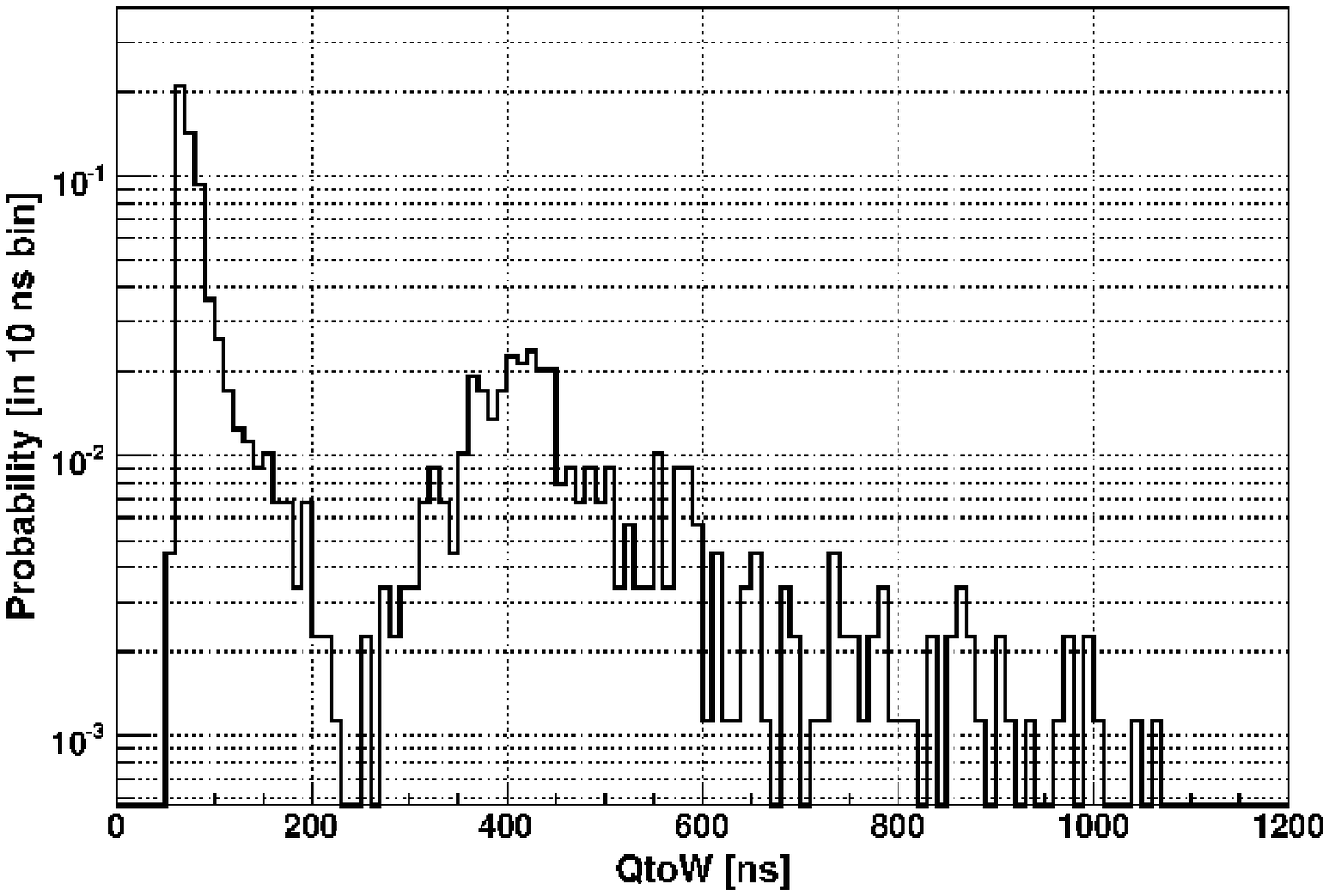}
\caption{\footnotesize Left: average behavior of the output signal width ($QtoW$) as a function of the input 
pulser charge together with a 4-parameter fit. 
The avalanche and streamer regions are indicated and also the threshold 
level. The dot-dashed line shows the
charge resolution resulting from the fit after including the fluctuations in the $QtoW$ value (right axis). 
Right: $QtoW$ distribution measured in the RPC under cosmic rays for an operating voltage $V=6.0$ kV. 
The two separate distributions can be attributed to avalanches and streamers.}
\label{Fig_TOT}
\end{center}
\end{figure}

The electron induced (prompt) charge $q_p$ was codified in the width of the FEE (LVDS) output signal
through a `charge to width' algorithm  \cite{Dani_HADES}, that will be referred as $QtoW$.
The algorithm is non-linear as illustrated in Fig. \ref{Fig_TOT} with avalanches lying mainly 
on a first (steep) linear part of the $QtoW$ vs $q_p$ correlation
curve, while streamers are concentrated in a second (soft) one. 
Despite the non-linearity, avalanches and streamers 
can be well resolved (Fig. \ref{Fig_TOT}-right) 
and appear separated at around $q_p\simeq 5$ pC ($QtoW=200$ ns).
This `QtoW method' can indeed accommodate a very large dynamic range 
while keeping the charge resolution below 10\%-$\sigma$ for avalanche-like pulses
with $q_p>50$ fC. 
(Fig. \ref{Fig_TOT}-left, dot-dashed line).
Points in Fig. \ref{Fig_TOT}-left were obtained by injecting
the charge with a fast square pulser ($t_{rise}=0.35$ ns) after
differentiation ($C=1$ nF, $R=50 \Omega$) in order to emulate the shape of the RPC signals.
Data was taken with a Tektronics TDS7104 Oscilloscope (BW=1GHz).

The bi-linear behavior of the average signal width ($\overline{QtoW}$) as a function of the average
prompt charge ($\bar{q}_{p}$) can be embodied in a simple parameterization as:
\beq
\overline{QtoW}=a(1- e^{-b \bar{q}_{p}}) + c \bar{q}_p + ~ QtoW_{min} \\ \label{cal_curve}
\eeq
after which the resolution of the 'QtoW method' can be obtained:
\bear
&& \frac{\sigma_{q_p}}{\bar{q}_p}=\frac{\partial \bar{q}_p}{\partial \overline{QtoW}}
~\frac{\sigma_{_{QtoW}}}{\bar{q}_p} \\
&& \frac{\sigma_{q_p}}{\bar{q}_p}=\frac{1}{a b e^{-b\bar{q}_p}+c}
~\frac{\sigma_{_{QtoW}}}{\bar{q}_p}
\eear
 $QtoW_{min}$ (53 ns for the channel shown in Fig. \ref{Fig_TOT}) is the minimum output 
signal width, corresponding to the
minimum time during which the comparator is being self-latched.
 By numeric integration of the signal amplitude, a close agreement for the
$QtoW$ vs $q_p$ curves was observed when analyzing avalanches 
originated inside the detector as compared with pulser data, while 
streamers clearly deviate from the pulser behavior (up to a factor 1/2 less in the reconstructed $q_p$).
The fluctuations of the signal width are of the order of $\sigma_{_{QtoW}}=200$ ps
 and no dependence with the input charge was observed.
A detailed description of the performances of the final HADES-FEE will
be the subject of a more technical forthcoming publication and will not follow in this paper.

Eq. \ref{cal_curve} provides an accurate phenomenological description and illustrates the different
behavior of the algorithm for low and high charges, but cannot be inverted, so in practice a $6^{th}$
order polynomial was used to obtain the `calibration curve' $q_p(QtoW)$. To avoid errors resulting 
from an incorrect extrapolation, values of the signal width in excess of 290 ns are considered as 
an `overflow' and the maximum pulser charge of
$q_p=25$ pC is assigned to them. Such big charges are rarely achieved under ordinary circumstances and 
they occurred seldom even in the very harsh environment studied here.

\subsection{Total charge and rate determination}

The dynamic behavior of an RPC at high rates is bound to the average total avalanche charge 
$\bar{q}_{_T}$ \cite{Aielli,Diego_B}. Being
readily obtainable by a direct current measurement (after dividing by the avalanche rate), its value
is directly related to the average gap voltage once the stationary (DC) situation is reached, as:
\beq
\bar{V} = V-\bar{I} R = V-\bar{q}_{_T} \Phi \rho d \label{DC}
\eeq
with $\Phi$ the avalanche flux [in Hz/cm$^2$], $\rho$ the electrode resistivity and $d$ its thickness. 
$V$ is the applied voltage, $\bar{I}$ the average current and $R$ the electrode 
resistance.
As compared with the stationary (DC) situation described by eq. \ref{DC}, the 
interpretation of $\bar{q}_{_T}(t)$ during the stabilization of the field
in the gap is more intricate \cite{Miguel_NIM}. Although it is not the main focus of the present
work to study the transient behavior of the RPC cells, proper means for determining the average avalanche
charge as a function of the irradiation time were devised as follows. The particle rate was measured 
from scintillator S$_1$ as the coincidence of signals from its two PMs.
This provides an unbiased rate estimate by suppressing single electron noise. The coincidence signal was sent
to the scaler input of a commercial LabJack U3 acquisition board connected to a computer,
and its value stored every 0.2 s. Complementary, 
the RPC current was measured via the analog output of a 2-channel CAEN N471A HV supply, after calibration.
Although the HV-display resolution is $\pm 1$ nA, by averaging the analog output the resolution could be 
improved
down to $\simeq 0.5$ nA. Due to the presence of environmental noise, the measured current
had to be averaged over a pretty large time interval of $\simeq 0.8$ s being stored every $0.2$ s, and read-out
with the same acquisition board. With the help of a second scaler input in the LabJack U3, 
the cycles of an external 40 kHz clock were counted in order to provide a stable time 
estimate.

\subsection{The physical environment}

\begin{figure}[ht!!!]
\begin{center}
\includegraphics[width = \linewidth]{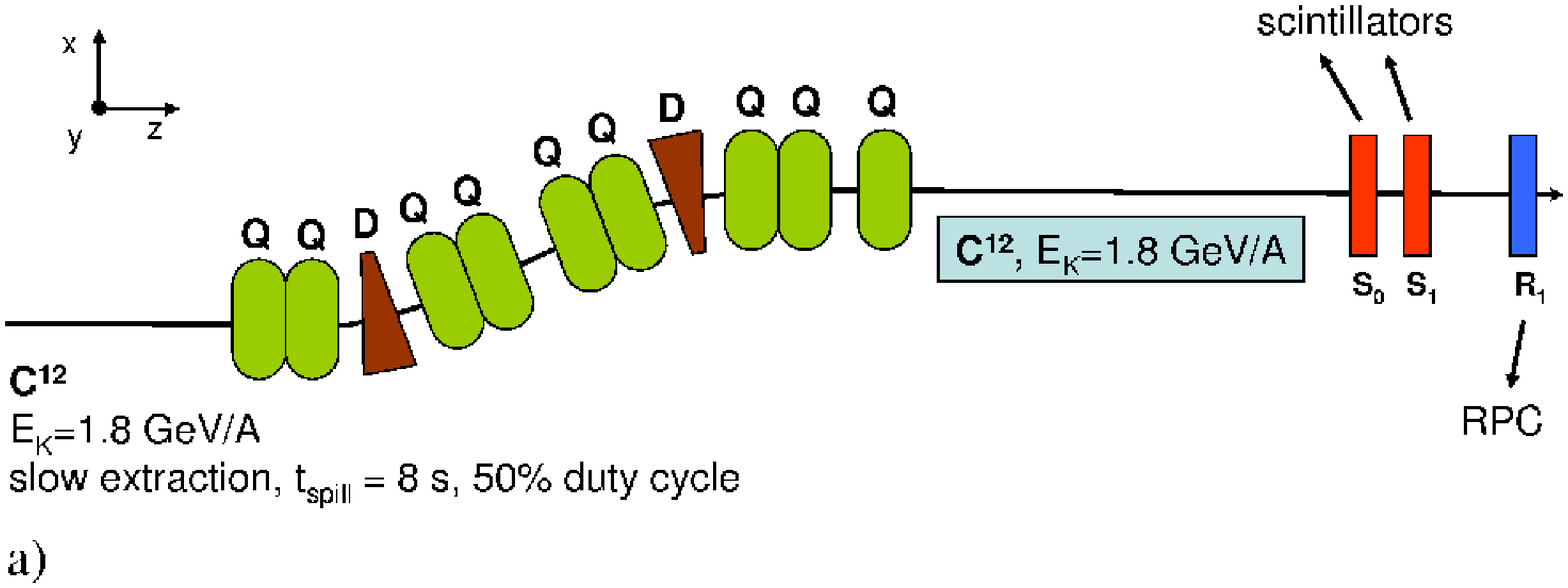}
\includegraphics[width = \linewidth]{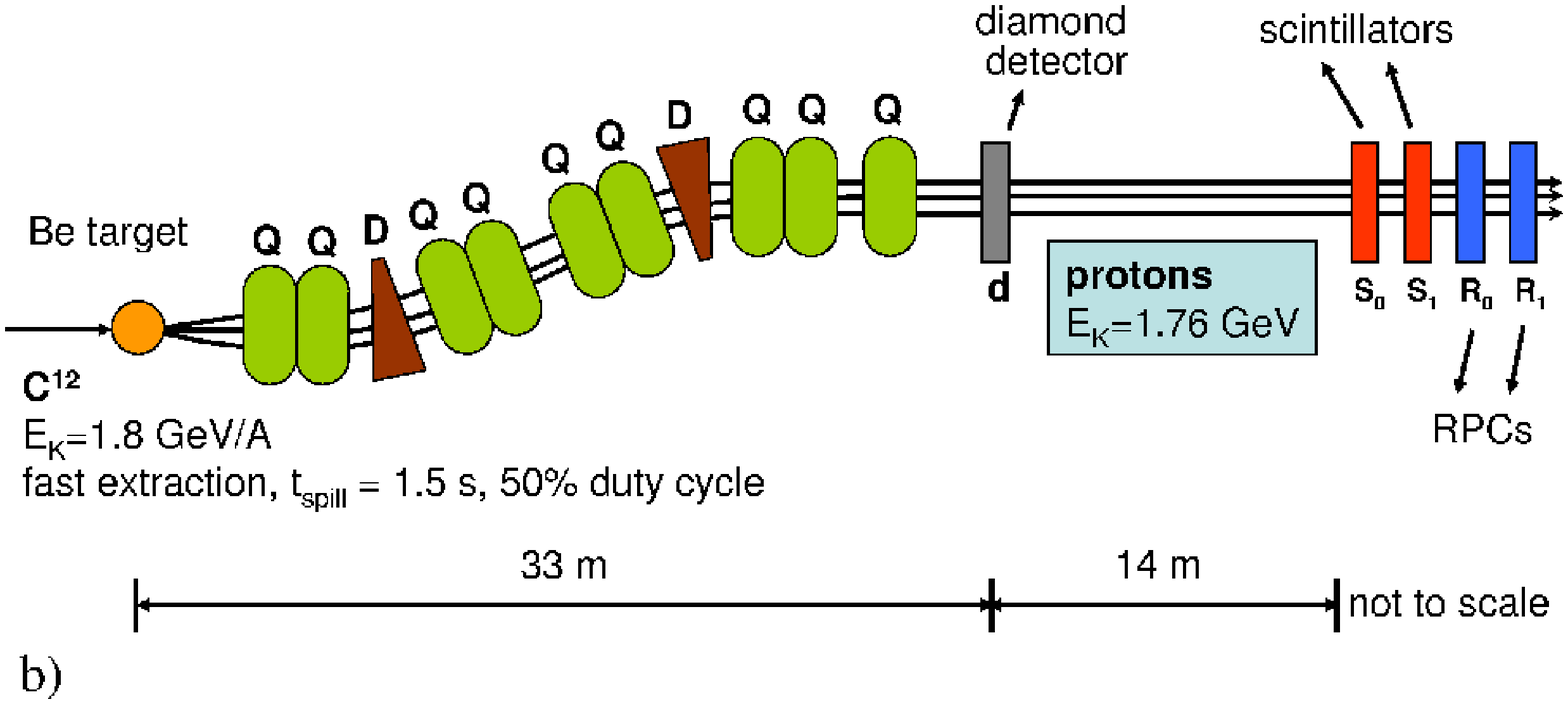}
\includegraphics[width = \linewidth]{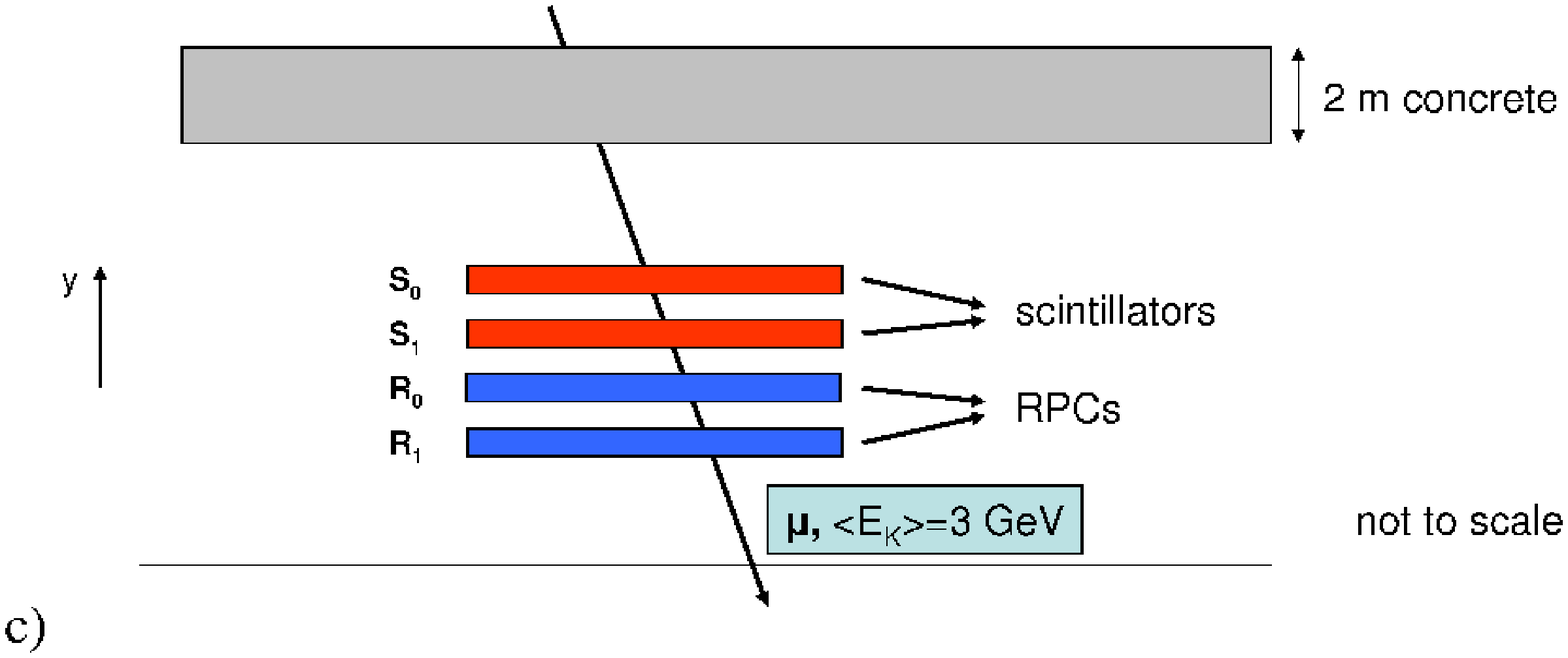}
\caption{\footnotesize a) Setup for focused C$^{12}$ irradiation 
(kinetic energy $E_K=1.8$ GeV/A, $\sigma_{_{E_K}}/E_K<1\%$). 
b) Setup for diffuse proton irradiation ($E_K=1.76$ GeV, $\sigma_{_{E_K}}/E_K=4\%$) after C$^{12}$ reactions
in a secondary Beryllium target placed 33 m upstream the experimental hall. Diamonds have
been added and also a second RPC (R$_o$). c) Setup for cosmic muon detection.}
\label{setups}
\end{center}
\end{figure}

\subsubsection{C$^{12}$ beam}

\begin{figure}[ht!!!]
\begin{center}
\includegraphics[width = 4.5 cm, angle=90]{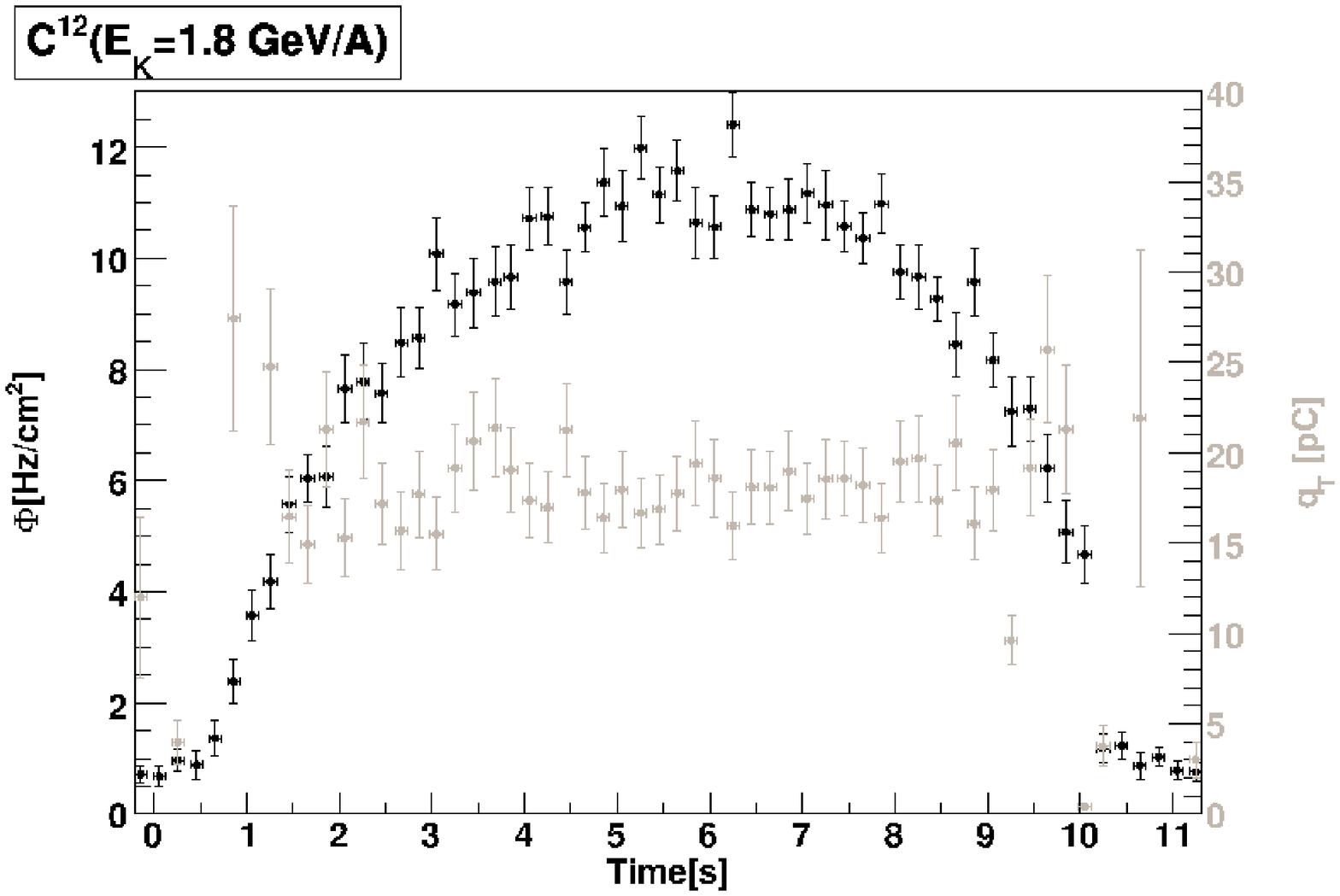}
\includegraphics[width = 4.5 cm, angle=90]{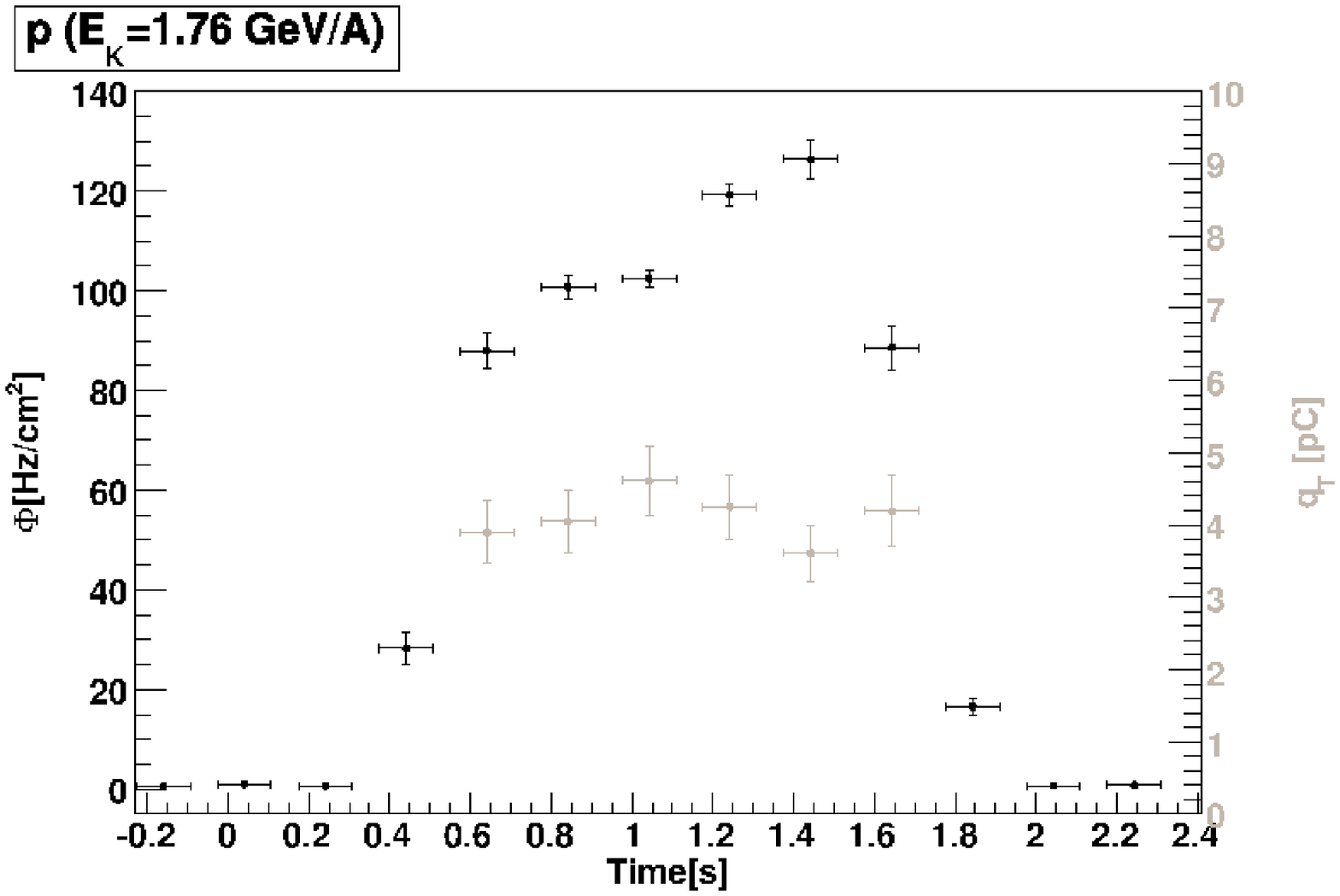}

\includegraphics[width = 4.5 cm, angle=90]{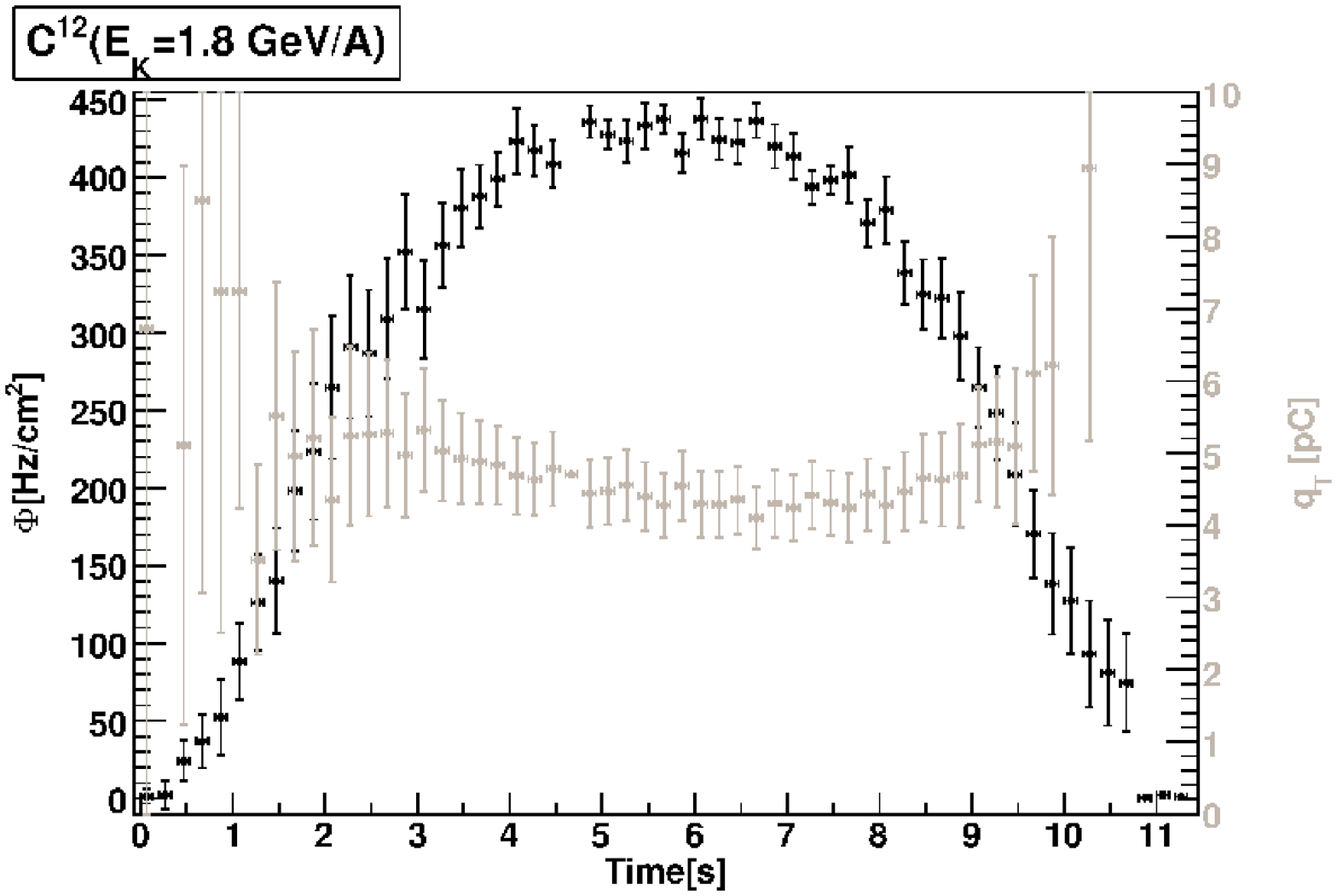}
\includegraphics[width = 4.5 cm, angle=90]{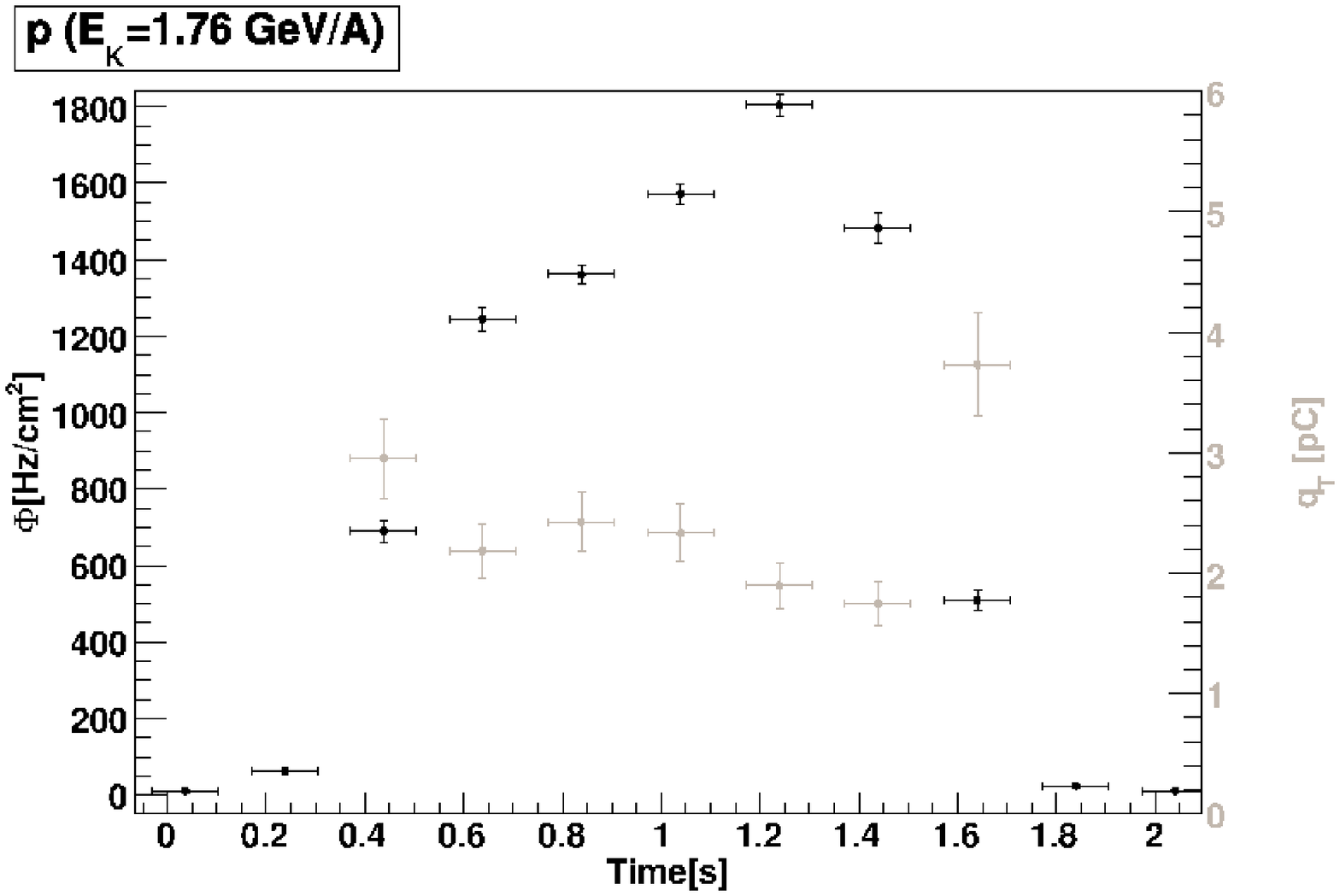}

\caption{\footnotesize Illustration of characteristic
spill time-profiles. The left column shows the average flux profile (over many spills)
and the average total charge $\bar{q}_{_T}$ behavior (right axis) for illumination
with C$^{12}$ (kinetic energy $E_K=1.8$ GeV/A) at two different beam intensities.
Right column shows similar plots
for diffuse proton illumination ($E_K=1.76$ $\pm 4\%$ GeV). The extraction time was about a factor 5 smaller in such a case.
The transient behavior is little visible since the measured current had to be averaged over a pretty large
time-interval of 0.8 s due to external noise.}
\label{time_profile}
\end{center}
\end{figure}

Measurements were performed with three distinct setups. In a first one (Fig. \ref{setups}a)
C$^{12}$ was injected in the experimental area at a kinetic energy $E_K=1.8$ GeV/A directly from SIS18.
The beam was 
focused at the usual HADES target position, 14 m upstream our reference detectors. The latter were placed
close to the beam dump and optically aligned with respect to the beam-line. A typical beam profile at the
level of few mm$^2$ was expected in the target transverse (XY) plane, from which experience tells
that the transverse dimension of the primary beam amounts to some cm$^2$ at the beam dump. The accelerator
was operated in the so-called slow extraction mode that allows for a fairly sustained beam intensity
along 8 seconds spill (effectively) with a duty cycle close to 50\%. Typical spill time-profiles are shown
in Fig. \ref{time_profile} (left).

\begin{figure}[ht!!!]
\begin{center}
\includegraphics[width = 6.7 cm]{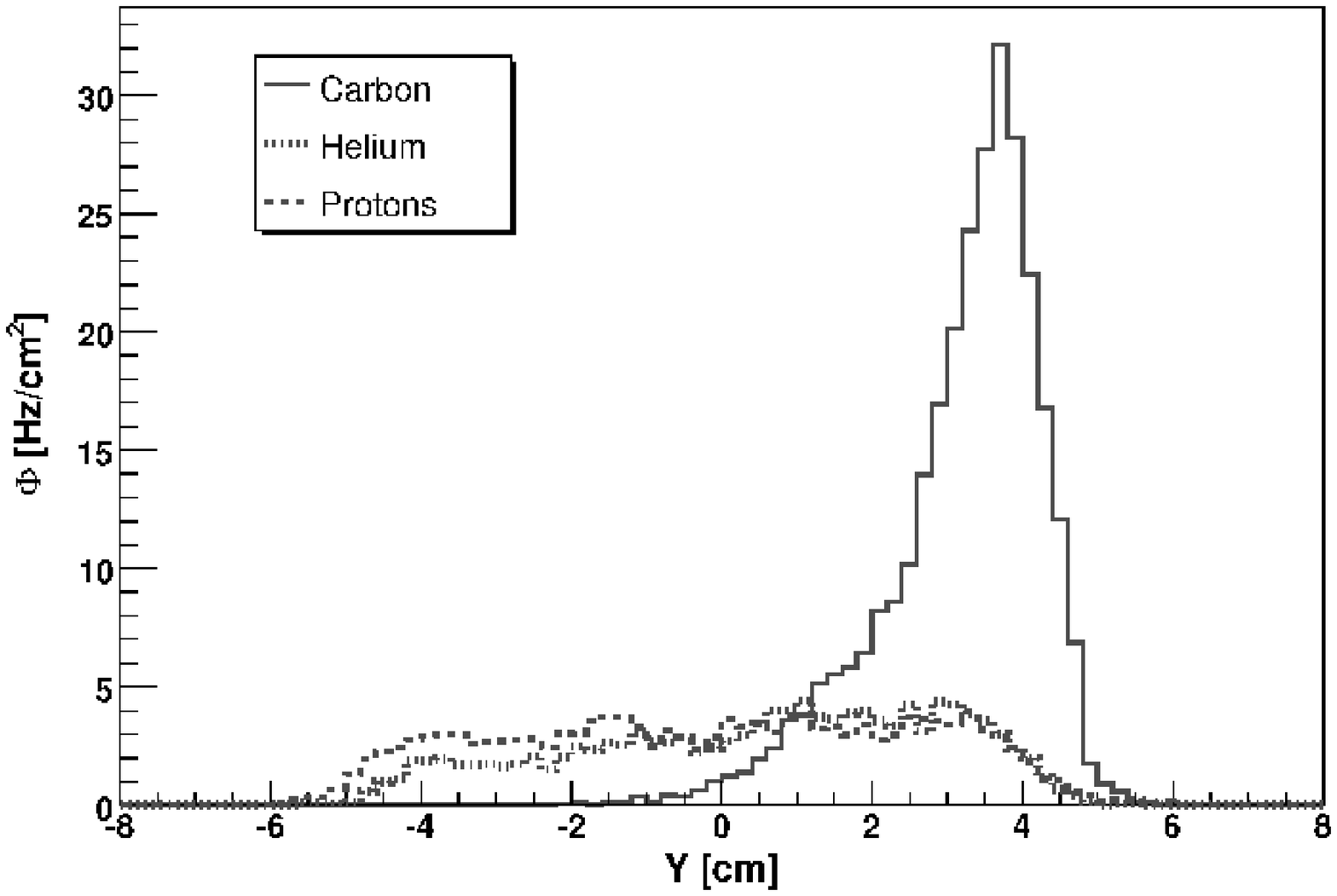}
\includegraphics[width = 7 cm]{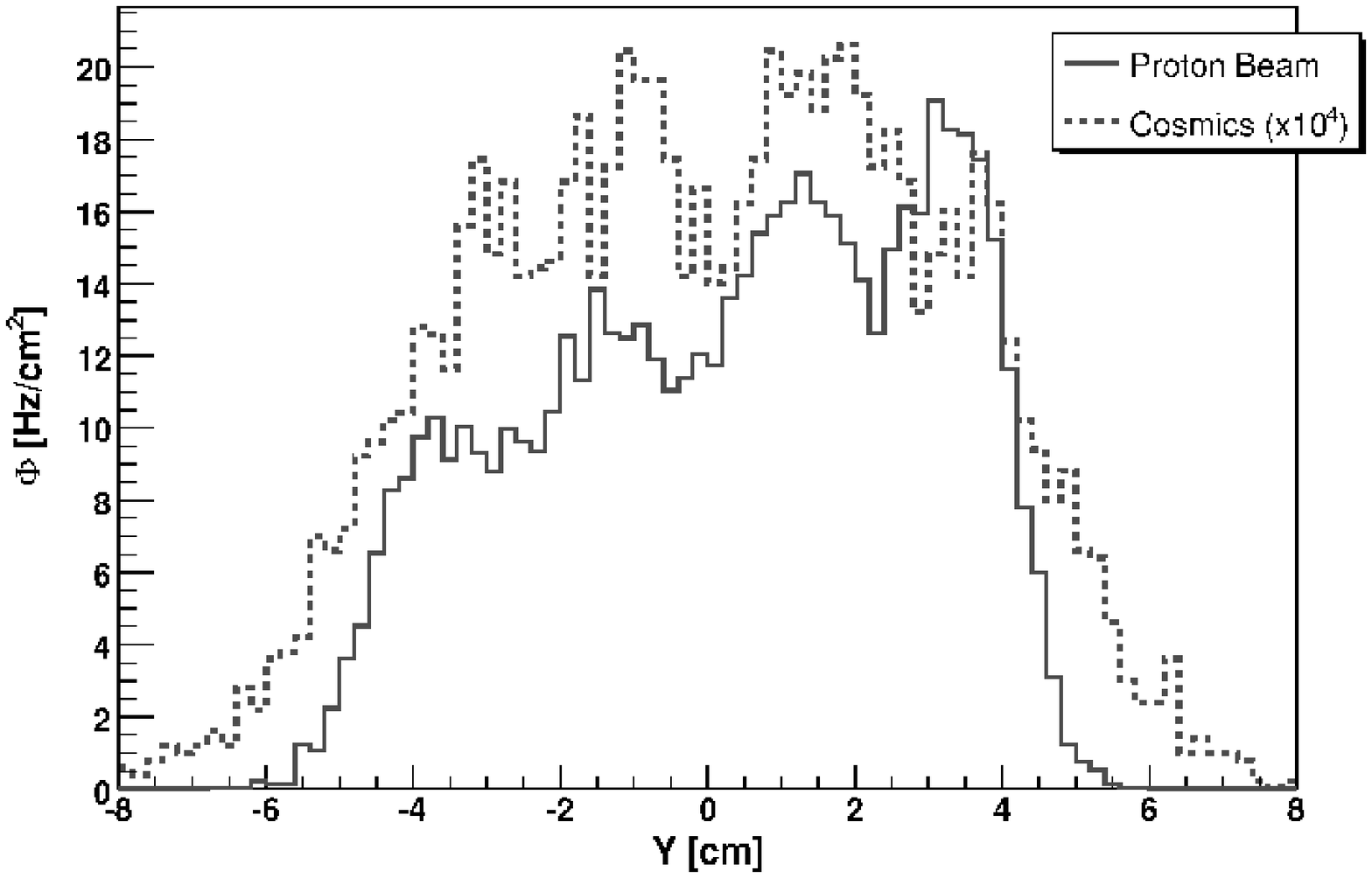}
\caption{\footnotesize Left: Position distribution along the reference scintillator $S_o$
under C$^{12}$ illumination. The proton, He and C$^{12}$ curves are shown after charge selection,
manifesting the primary C$^{12}$ beam (kinetic energy $E_K=1.8$ GeV/A) 
close to the border of the reference counters. Right: Position distribution
under diffuse proton illumination  ($E_K=1.76$ $\pm 4\%$ GeV) and cosmic ray profile (overlaid). 
The slightly bigger profile under cosmic rays stems from the larger angular spread (see text).}
\label{position_profile}
\end{center}
\end{figure}

\begin{figure}[ht!!!]
\begin{center}
\includegraphics[width = \linewidth]{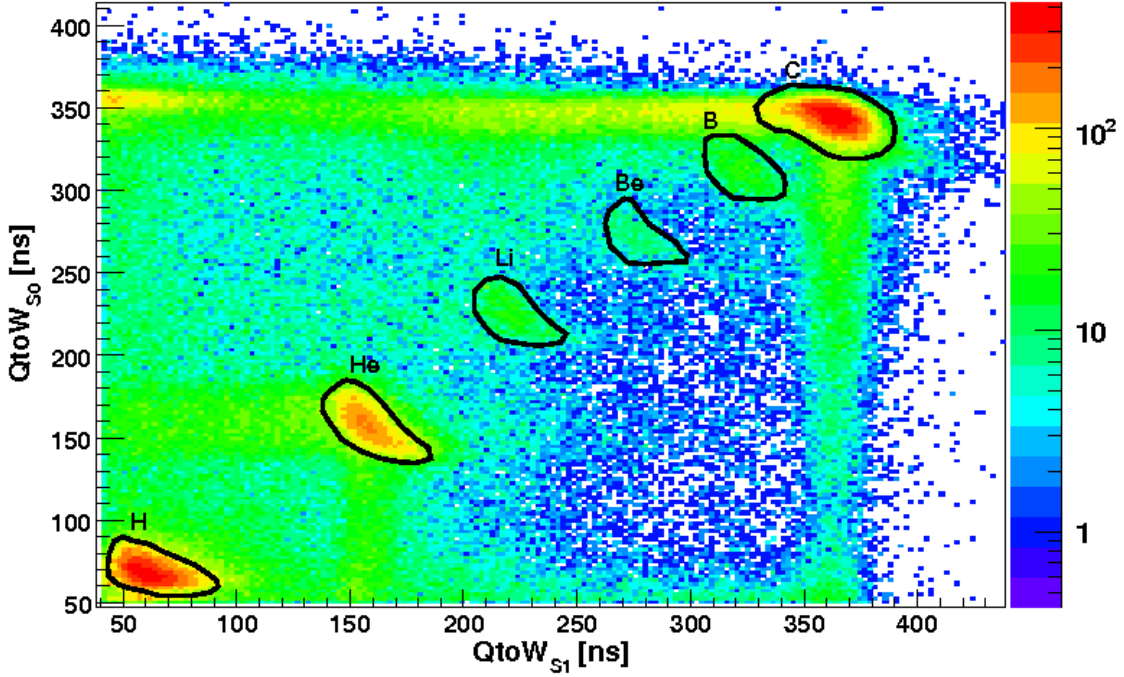}
\caption{\footnotesize Two-dimensional correlation plot between the raw signal widths obtained
 in the reference scintillators after offset subtraction (color on-line). Six 
blobs can be identified being the most prominent ones attributable to protons, He and C$^{12}$. The other
3 blobs with a much lower yield must be Li, Be and B or charge states of the C$^{12}$ ion.}
\label{scin_cuts}
\end{center}
\end{figure}

The non-perfect detector-beam
alignment resulted in a position distribution over the reference scintillator system
characterized by a narrow peak ($1$ cm$-\sigma$) close to one of the detectors ends, submersed 
in an uniform background  (Fig. \ref{position_profile}-left). After selection based on the signal
width in the reference scintillators (Fig. \ref{scin_cuts}), the peak could be attributed
to the primary C$^{12}$ beam while the uniform
background was largely populated by secondary protons and He together with species with $Z=3, 4, 5$
(either Carbon charge states or Li, Be, B species), the latter at a much lower yield. Particles different
from Carbon must have been originated along 
the roughly 14 meters downstream from the exit of the vacuum pipe to the
detector setup. Due to the different shape of the scintillator signals as compared with RPC
avalanches, the absence of amplifier and the eventual saturation of the PMs at the higher charges, 
the calibration curve from Fig. \ref{Fig_TOT} could not be used and we present in Fig. \ref{scin_cuts}
the raw values of the signal width ($QtoW$). Based on them we believe, nevertheless, that identification
can be performed with little ambiguity.

The position in
the scintillators was determined by constraining the width of the time difference distribution (left-right)
to the known detector dimensions. In such a case the propagation velocity can be obtained as:
\beq
v_{prop} = 2 \frac{L}{(t_{L}-t_{R})}_{50}
\eeq
where ${(t_{L}-t_{R})}_{50}$ refers to the width of the time difference
profile at 50\% drop from the flat top. The effective light propagation velocities obtained were $v_{prop,s_o} = 0.385 c$
and $v_{prop,s_1} = 0.380 c$, being $c$ the speed of light.
Finally, the detectors were mutually aligned by software by imposing that the average position 
difference ($\bar{y}_{s_o}-\bar{y}_{s_1}$) was centered at zero. 

Only one RPC cell ($R_1$) was used during these measurements since $R_o$ could not be powered
up due to technical reasons. A constant voltage $V=5.6$ kV was applied to the RPC, 
while the primary flux ranged from 10 to 1000 Hz/cm$^2$.

\subsubsection{Proton beam}
\label{proton_beam}

In a second setup (Fig. \ref{setups}b), C$^{12}$ at $E_K=1.8$ GeV/A was collided into
a Beryllium secondary target placed 33 m upstream the experimental area. The magnetic field of the following
dipoles and quadrupoles in the beam line was adjusted for accepting forward protons from
the reaction at an average energy $E_K=1.76$ GeV (that accounts for energy losses in
the Be target itself). As a consequence of Fermi motion, protons had still a sizeable angular straggling  
that rendered a much more uniform irradiation than for the C$^{12}$ beam (Fig. \ref{position_profile}-right).

In this setup, auxiliary mono-crystalline diamonds were placed 14 m upstream the scintillator
reference system.
From the time spread between the diamond detector and the first scintillator,
$\sigma_{_{T(D-S_o)}}=248$ ps, and the simultaneously measured scintillator and
diamond resolutions ($\sigma_{_{T(D)}}=150$ ps, $\sigma_{_{T(S_o)}}=\sigma_{_{T(S_1)}} = 40$ ps)
the energy spread of the proton beam after propagation over $D = 14$ m could be determined as:
\beq
\frac{\sigma_{E_K}}{E_K}=\frac{m_p}{E_K}\frac{\beta^2}{^{3/2}\sqrt{1-\beta^2}}\frac{\sigma_{tof}}{\tn{tof}}
\label{spread}
\eeq
being
\bear
&& \tn{tof}     = \frac{D}{\beta c} \\
&& \sigma_{tof} = \sqrt{\sigma_{_{T(D-S_o)}}^2-\sigma_{_{T(D)}}^2-\sigma_{_{T(S_o)}}^2}
\eear
resulting in a value for $\frac{\sigma_{E_K}}{E_K}=4\%$. The energy losses
due to the presence of 14 m of air are much below $4\%$ for protons at this energy, supporting
that the energy straggling is indeed originated
in the production at the Be target. Although the time resolution of the diamonds
was reduced further down to the level of $100$ ps after a special analysis
\cite{Wolfgang}, this discussion remains out of the scope of the present paper.

The accelerator was also operated in slow extraction mode but at a reduced
extraction time (1.5 seconds), resulting in a slightly asymmetric spill time-profile
(Fig. \ref{time_profile}-right). 
Two RPCs (R$_o$, R$_1$) were powered up and measurements were taken as
a function of the particle flux in the range $10$-$1000$ Hz/cm$^2$ at voltages from $5.0$-$5.8$ kV.

\subsubsection{Cosmic rays}
\label{cosmic_rays}

We took data, after the measurements
of the previous two sections were performed, in a `reference setup' with a cosmic stand
where the scintillators and RPC cells were assembled according
to the previous section but horizontal with respect to the ground.
From the 2 meters of concrete that cover the HADES experimental area it can
be estimated that the main source of particles traversing the 4 detectors are muons at an average
energy $\bar{E}_{K}=3$ GeV ($\simeq 1$GeV energy loss) \cite{PDG}. The position profile obtained from
scintillator $S_o$ as shown in Fig. \ref{position_profile}-right is fairly uniform but a bit broader
than for proton irradiation, due to the larger angular straggling 
(since $S_1$ is a bit smaller than $S_o$ and is also in trigger, for
perpendicular incidence the measured $S_o$ length is very close to the $S_1$ one and therefore
$\simeq 2$ cm smaller than its true length). The HV was varied in the range $5.6$-$6.0$ kV.



\section{Results}
\label{results}
\subsection{Performance of the reference system}

\begin{figure}[ht!!!]
\begin{center}
\includegraphics[width = 13 cm]{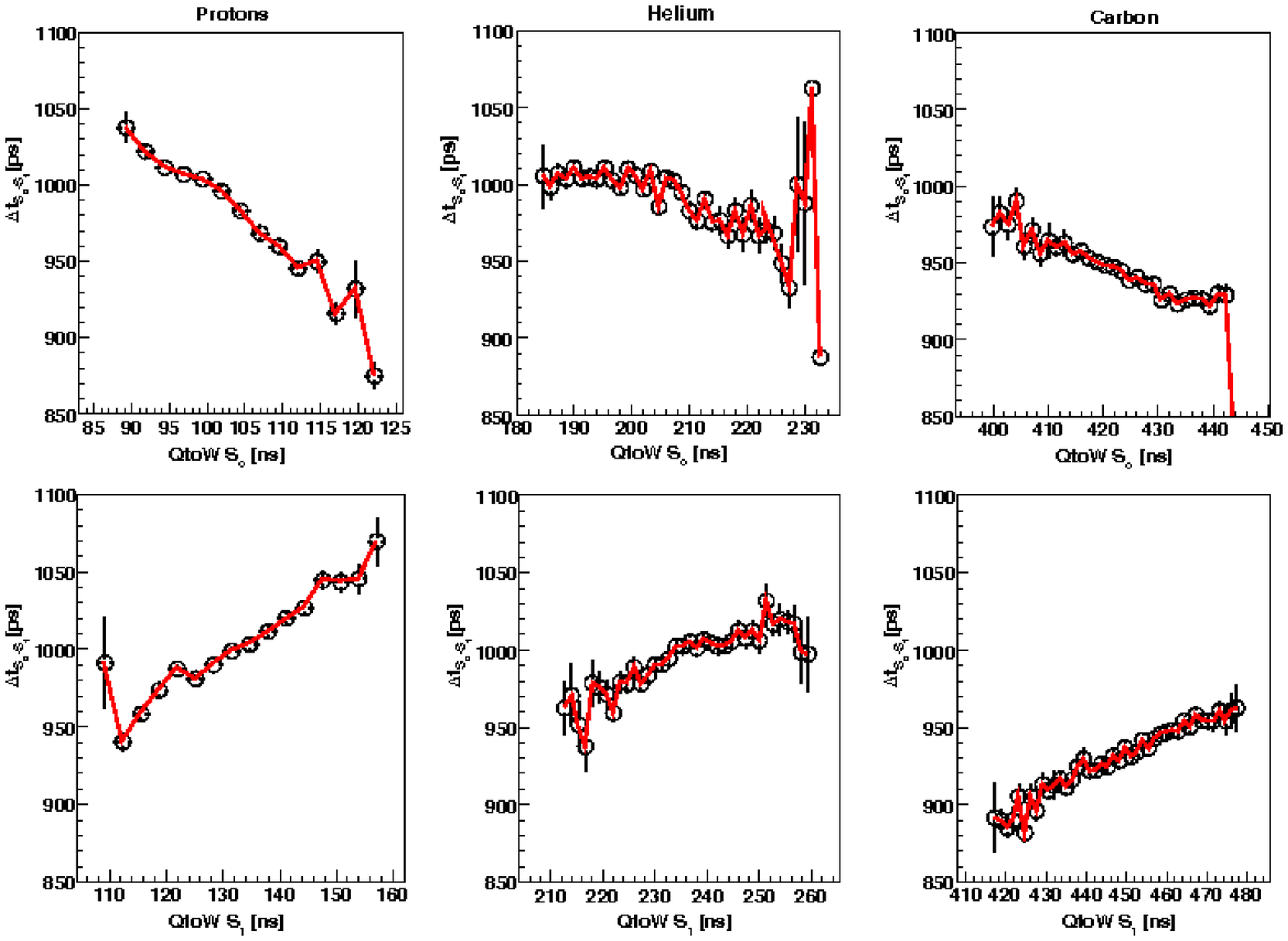}
\includegraphics[width = 12 cm]{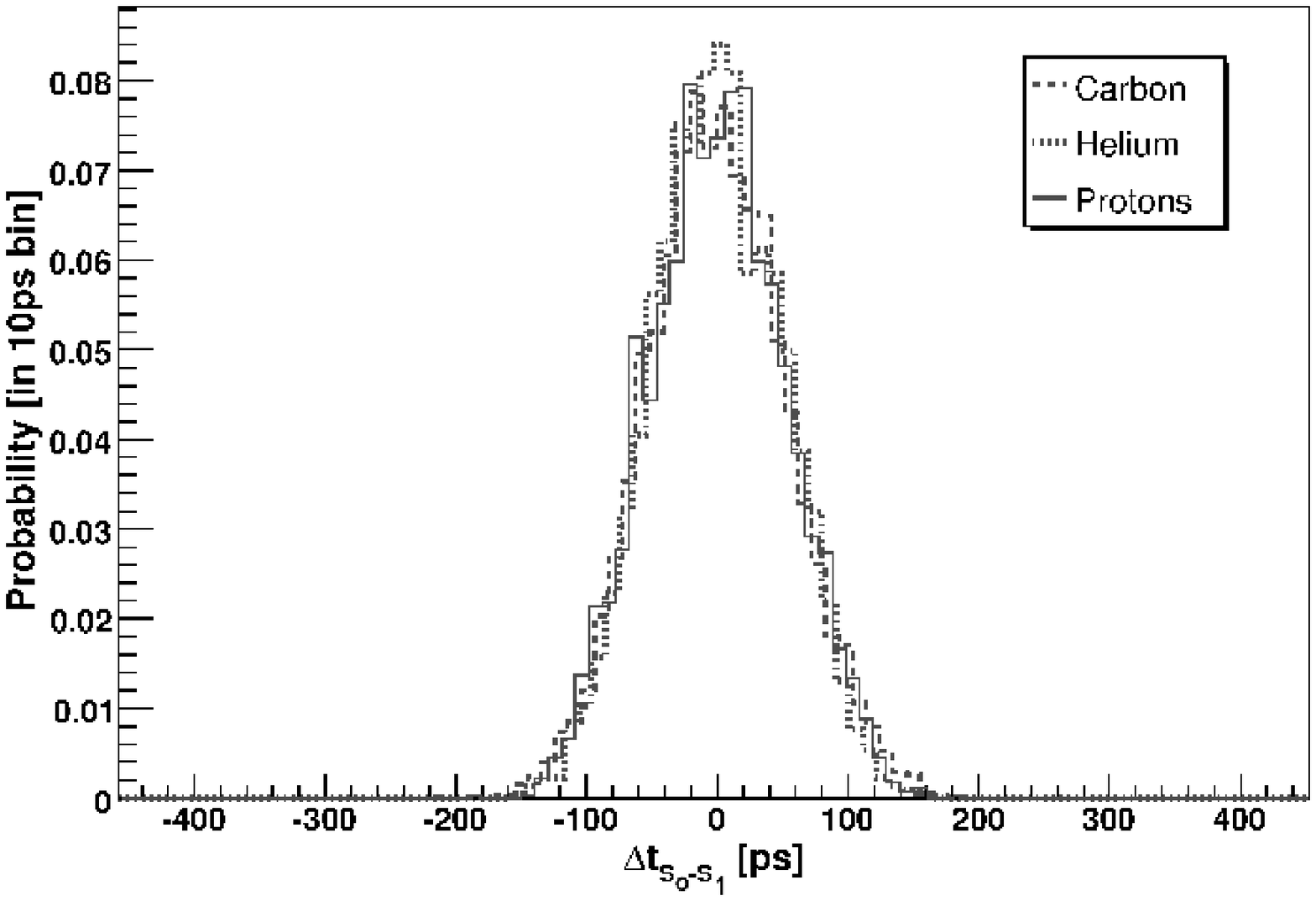}

\caption{\footnotesize Up: multi-linear fit used for correcting the walk of the
time of flight between the reference scintillators ($\Delta{t}_{S_o-S_1}$) 
as a function of the signal width ($QtoW$) for p, He and C$^{12}$. 
Down: distribution of times of flight between scintillators after the correction procedure 
for the three species.}
\label{carga_uncorr}
\end{center}
\end{figure}

The use of LED discriminators for the reference scintillators requires of time-charge slewing
corrections in order to obtain the best timing.
The correlation of the time-of-flight between $S_o$ and $S_1$ ($\Delta{t}_{S_o-S_1}$)
with the average signal width is shown in Fig. \ref{carga_uncorr}, being:
\beq
\Delta{t}_{S_o-S_1}= \frac{t_L+t_R}{2}\Bigg|_{S_o} - \frac{t_L+t_R}{2}\Bigg|_{S_1}
\eeq
The different species from $Z$=$1$-$6$ could be cleanly separated as shown in Fig. \ref{scin_cuts}
and a time-width correction was performed individually for each specie on the basis of multi-linear
segments.
After this procedure, a combined time resolution of $\simeq55$ ps was obtained for $\Delta{t}_{S_o-S_1}$
that, assuming both detectors to perform equally, yielded per counter $\sigma_{_T}=40$ ps for C$^{12}$,
$\sigma_{_T}=39$ ps and $\sigma_{_T}=42$ ps for secondary He and p.
The final time-of-flight distributions after corrections are shown in Fig. \ref{carga_uncorr}
for different particle species, showing a strong Gaussian behavior with $\sigma_{_T}\simeq40$ ps per counter,
while the position resolution obtained from the position difference between the counters
was $\sigma_{y}=6$ mm. No dependence with the particle specie was observed. Being
the values for $\sigma_{_T}$ consistent with the electronic resolution of the mean-time obtained with pulsed
signals \cite{Alex_LV} and not depending on the primary ionization, provides a strong evidence that the
performance of the reference system was indeed limited by the TDC resolution.

\subsection{RPC performance}

\begin{figure}[ht!!!]
\begin{center}
\includegraphics[width = 11 cm]{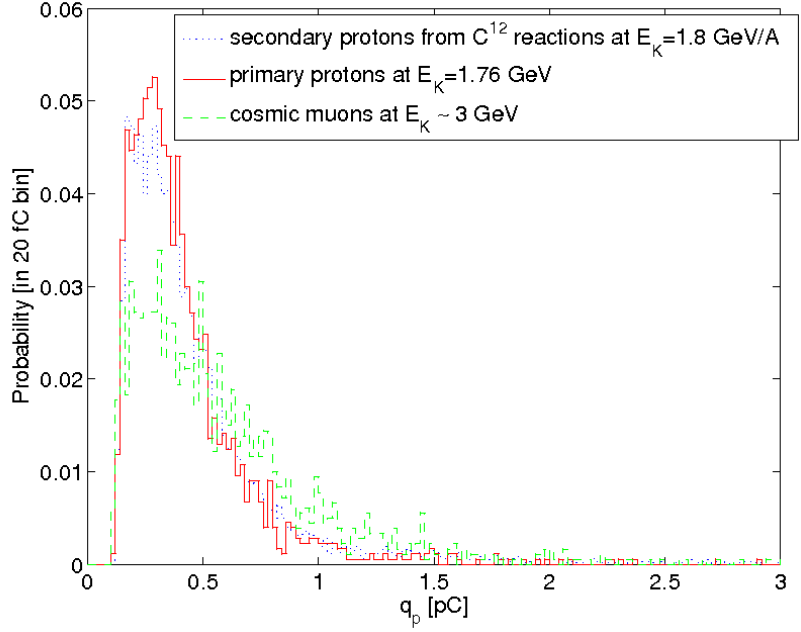}
\caption{\footnotesize Comparison of the charge distribution measured in 
the RPC under a diffuse proton beam at a kinetic energy $E_K=1.76$ GeV $\pm 4\%$ 
(continuous line), secondary forward protons 
from C$^{12}$ interactions at $E_K\sim1.8$ GeV (dotted) and cosmic muons at $E_K\sim3$ GeV (dashed).}
\label{charge_spectrum_p}
\end{center}
\end{figure}

The track selection was based on graphical cuts on the 2-dimensional signal width distribution
in the scintillators (Fig. \ref{scin_cuts}) for selecting the particle species corresponding
to $Z$=$1$-$6$. Due to
the energy regime, either secondary forward protons from C$^{12}$ up-stream reactions or primary ones with
$E_K\simeq 1.76$ GeV can be highly regarded as mips ($\gamma\beta \simeq 2.7$), 
the latter having additionally
very little energy spread. On the other hand,
an estimate of the energy loss from the Bethe-Bloch formula (eq. \ref{BetheBloch}) 
indicates that cosmic muons with $E_K\simeq 3$ GeV ($\gamma\beta \simeq 30$) can ionize $30\%$ more in the RPC gas
than mips. A normalized charge distribution after cuts is shown in
Fig. \ref{charge_spectrum_p} where indeed a very small difference is seen between the 3 different cases, 
being the average charge released by cosmic muons a bit higher, indeed, than that for mips.

\begin{figure}[ht!!!]
\begin{center}
\includegraphics[width = \linewidth]{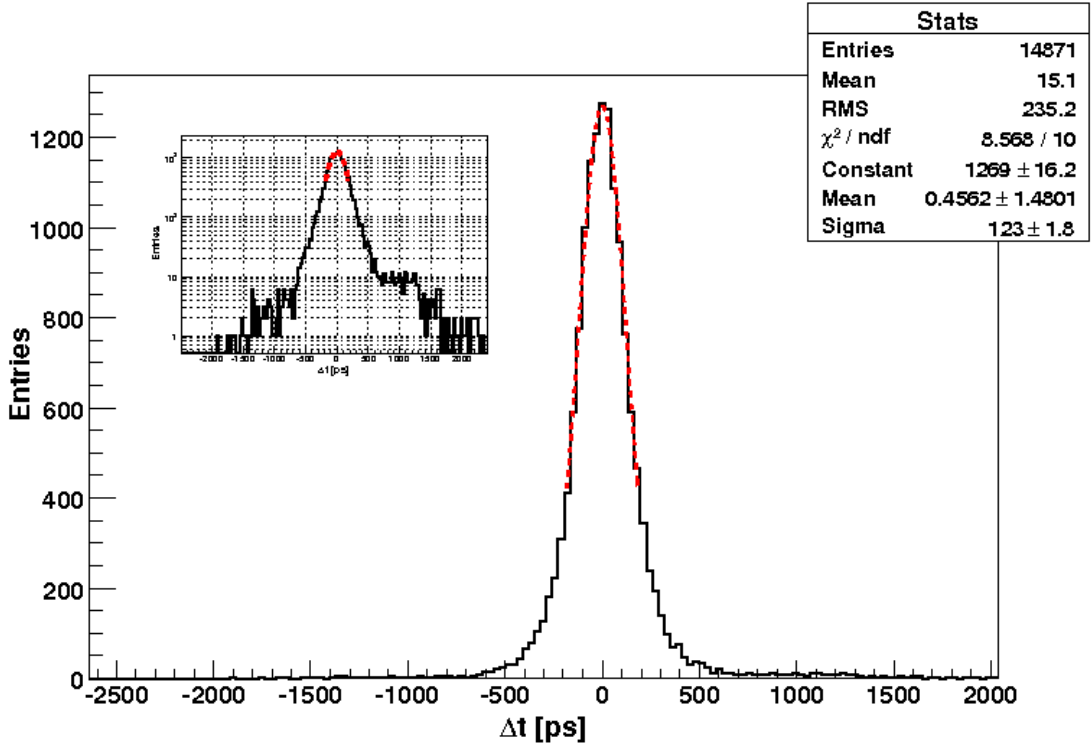}
\caption{\footnotesize Time of flight distribution between RPCs for a diffuse proton beam 
(kinetic energy $E_K=1.76$ GeV $\pm 4\%$) at 150 Hz/cm$^2$ and V=5.6 kV. The inset shows the 
same distribution in logarithmic scale, indicating the presence of tails at $1\%$ level. The 
counter resolution from the fit is $\sigma_{_T}=123/\sqrt{2}=87$ps.}
\label{resolution_RPC}
\end{center}
\end{figure}

After particle identification, the last 1 cm from the border of the scintillators was disregarded.
Furthermore, a `co-linearity' cut of $\pm 2$-$\sigma$ in the distribution of the position differences
between scintillators ($y_{s_o}-y_{s_1}$) was applied for improving the track quality
selection. In order to reduce the effect of the non-uniform irradiation, C$^{12}$ ions were additionally 
selected by a cut at $y>2$ cm and other species by $y<0$ cm.
After these cuts for enhancing the purity and quality of the reference tracks, the time of flight
between the two overlapping RPCs ($\Delta{t}_{R_o-R_1}$) was defined:
\beq
\Delta{t}_{R_o-R_1}= \frac{t_L+t_R}{2}\Bigg|_{R_o} - \frac{t_L+t_R}{2}\Bigg|_{R_1}
\eeq
and the time resolution per counter was obtained as
$\sigma_{_{T}} = \sigma_{\Delta{t}_{R_o-R_1}}/\sqrt{2}$.
Correction curves with respect to the charge of both RPCs (slewing correction) and the position
as given by the scintillators were performed. This is discussed in detail in section \ref{t-q-corr}.
 A typical time-of-flight distribution for protons at an uniform flux of $150$ Hz/cm$^2$
is shown in Fig. \ref{resolution_RPC}, yielding $\sigma_{_{T}}=87$ ps per RPC. We applied the usual
recursive fit around $\pm$1.5-$\sigma$ to reduce the influence of tails in the fit \cite{large}, 
but for this set of measurements the value of $\sigma_{_T}$ changed little indeed if a direct fit 
was performed.

In the C$^{12}$ experiment, due to the absence of the RPC cell $R_o$, the time difference with respect to
scintillator $S_o$ was calculated in a similar manner:
\beq
\Delta{t}_{R_1-S_o}= \frac{t_L+t_R}{2}\Bigg|_{R_1} - \frac{t_L+t_R}{2}\Bigg|_{S_o} \label{deltaS0}
\eeq
and the RPC resolution obtained after subtracting the contribution of the reference system:
$\sigma_{_T}=\sqrt{\sigma_{\Delta{t}_{R_1-S_o}}^2 - \sigma_{\Delta{t}_{S_o-S_1}}^2/2}$.

\subsubsection{Charge distribution}
\label{Charge_dist}

\begin{figure}[ht!!!]
\begin{center}
\includegraphics[width = 6.5 cm]{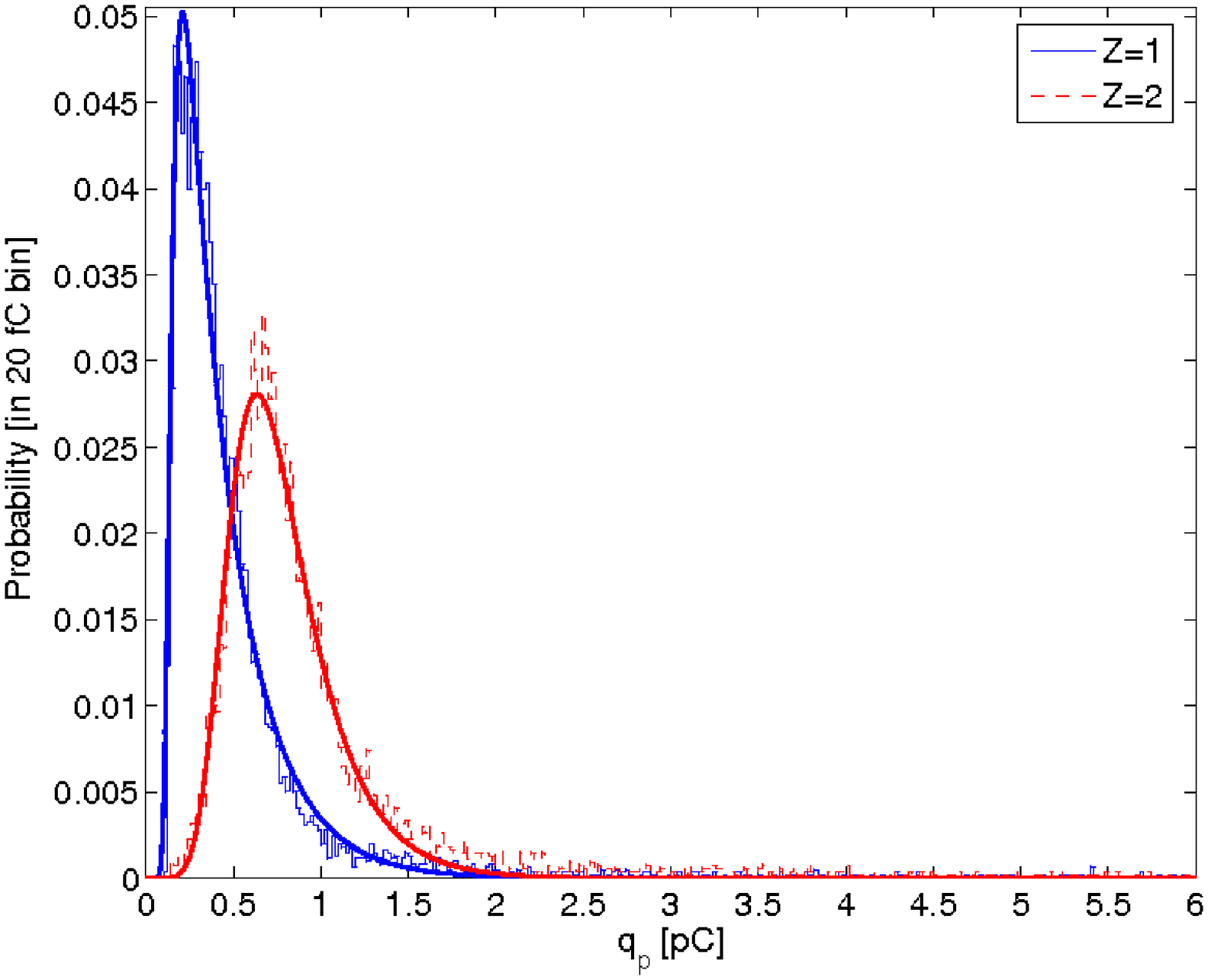}
\includegraphics[width = 6.6 cm]{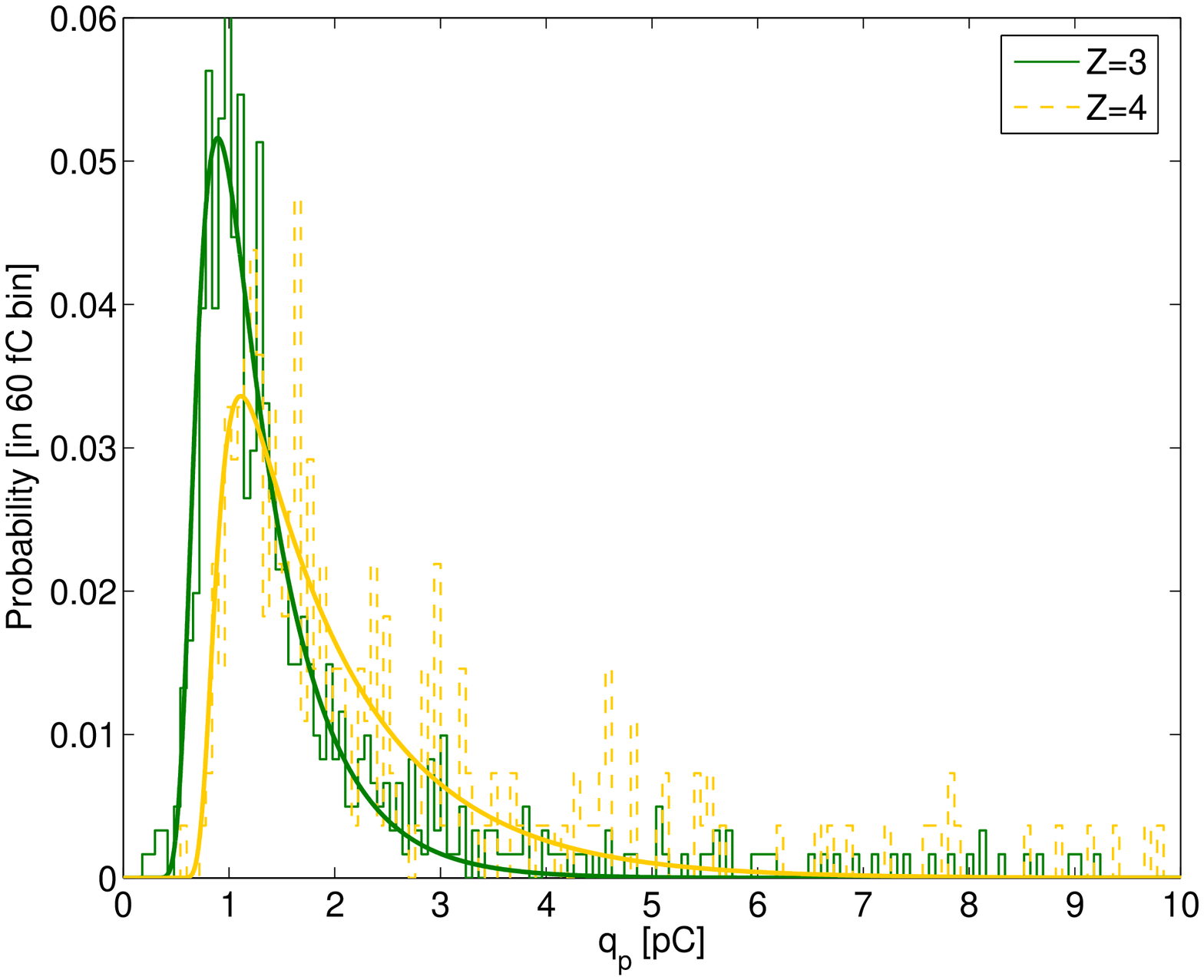}

\includegraphics[width = 6.5 cm]{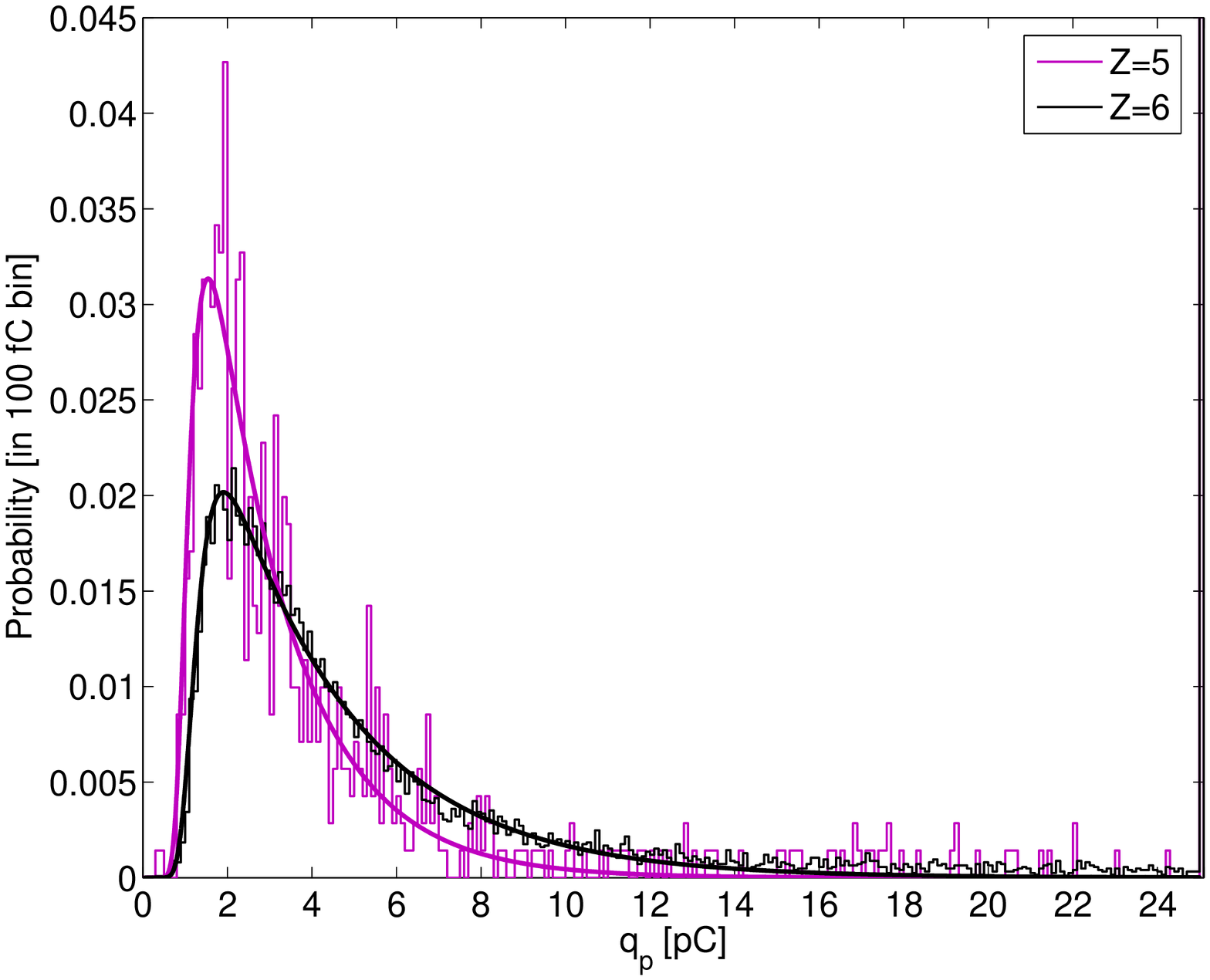}
\includegraphics[width = 6.5 cm]{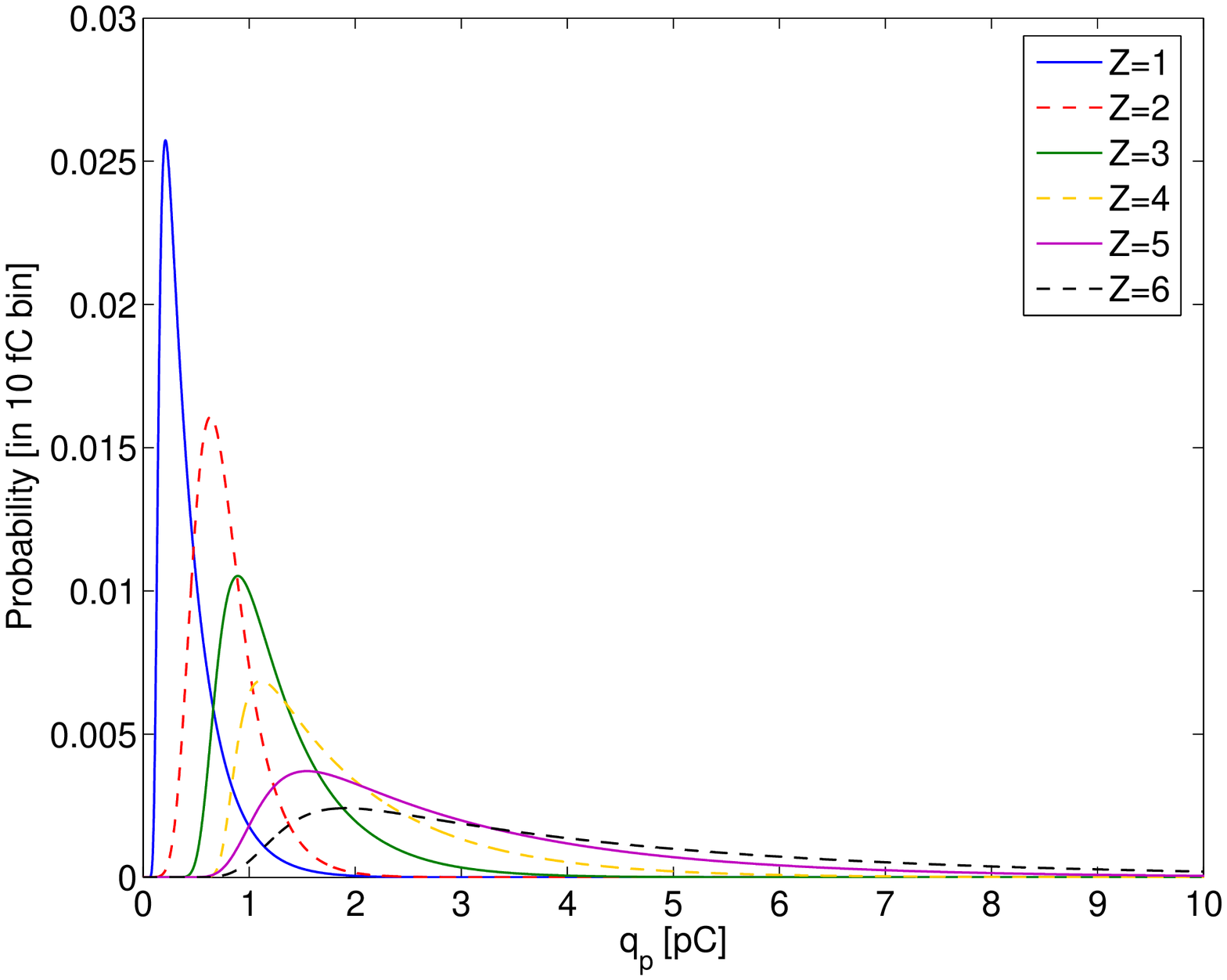}
\caption{\footnotesize Up-left: RPC charge distributions for fully 
stripped ions with atomic charges $Z=1,2$ together 
with a fit to a Landau-like function. Up-right: $Z=3,4$. Down-left: $Z=5,6$. Down-right: Fitted curves for
$Z=1$-$6$. The normalized momenta is $\gamma\beta\simeq2.7$ in all cases, V=5.6 kV and $\phi<10$ Hz/cm$^2$.}
\label{charge_spectrum_C}
\end{center}
\end{figure}

Normalized charge distributions are shown in Fig. \ref{charge_spectrum_C}
for $Z=1$-$6$ at $V=5.6$ kV at low rates ($\phi<10$ Hz/cm$^2$). A phenomenological fit 
to 4-parameter Landau-like distributions
\beq
\mathcal{P} = a \exp(-b q_p - c e^{-d q_p}) \label{Landau}
\eeq
provides a good description and partially captures the high charge tails. Fig. \ref{charge_spectrum_C}
shows that there is indeed a strong correlation between the charge initially released and the final
avalanche charge, but clearly not proportional, as one would expect if Space-Charge would be strong.
Nevertheless, specific experimental situations where protons must be separated from Carbon, for instance,
 can be certainly addressed with these counters. These issues are further discussed in section
\ref{Z_dependence}. Remarkably, it seems not to exist a clear separation between avalanches and 
streamers at high initial ionizations (contrary to the case when increasing HV, 
see Fig. \ref{Fig_TOT}). This fact
cannot be firmly attributed to the RPC dynamics since our charge calibration is sensitive to the signal 
shape and is therefore different for streamers, as pointed out in section \ref{Q_cal}. 
Due to that, the region $q_p>5$ pC can be probably not assessed precisely in this work.

\subsubsection{Dependence with HV}

\begin{figure}[ht!!!]
\begin{center}
\includegraphics[width = 12 cm]{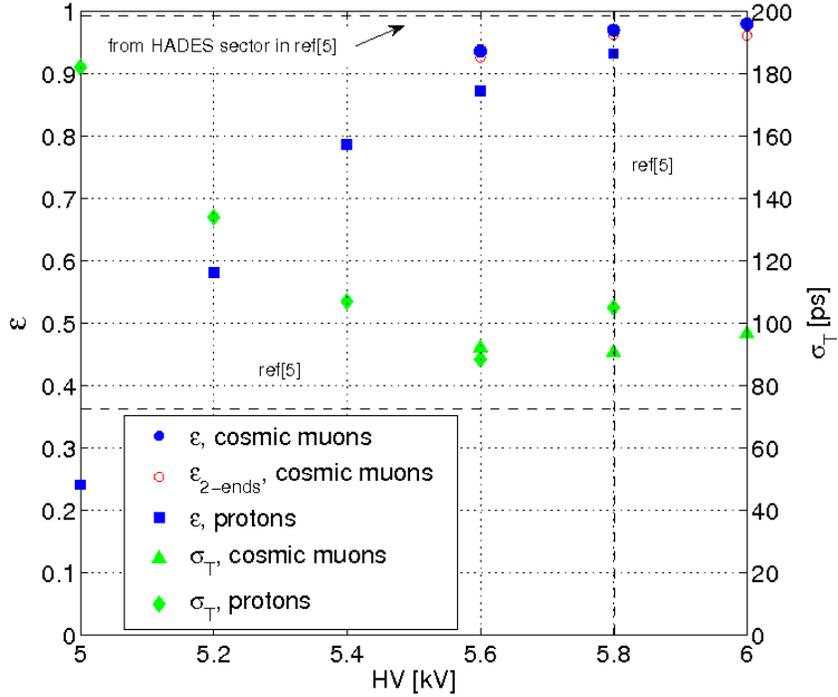}
\caption{\footnotesize Efficiency for cosmic muons as a function of High Voltage
when requiring a valid hit in least one detector end (full circles)
or in both (open circles), together with the efficiency for $E_K=1.76 \pm 4\%$ GeV protons (squares). 
Time resolution for cosmic muons (triangles)
and $E_K=1.76 \pm 4\%$ GeV
protons (diamonds) is also shown with scale on the right axis. A comparison with the system values
from \cite{Paulo_HADES,Alberto} (efficiency, resolution and working point) is indicated by dashed lines.}
\label{HV_scan}
\end{center}
\end{figure}

A scan in HV was performed for $E_K=1.76$ GeV diffuse protons and cosmic muon irradiation. A particle flux
of 150 Hz/cm$^2$ was chosen for the former while a much lower one was observed for the latter.
The behavior of the time resolution and efficiency is shown in Fig. \ref{HV_scan}. Closed an open circles
represent the efficiency when at least one or when both detector ends collect a valid signal, respectively,
showing little difference.
As compared with \cite{Alberto} the behavior is slightly (but consistently) worst.
While the resolution at the plateau is $\sigma_{_T}\simeq 90$ ps both for protons (diamonds) and cosmic muons
(triangles), and the efficiency $\varepsilon=92\%$ (squares) and $\varepsilon=96\%$ (circles) correspondingly,
the values reported for the HADES in-beam test \cite{Alberto} were $\sigma_{_T}=73$ ps,
$\varepsilon=99\%$ (intrinsic) and $\varepsilon=98\%$ (system). Nevertheless, in such a case,
no particle identification was available and actually a fraction of the
impinging particles is expected to be more ionizing than mips. This, together with the slight redundancy of the
HADES system ($\simeq 30\%$ cell overlap) may explain the slightly worse efficiencies observed here for mips.
The worse resolution may be
attributed to the grounding scheme and overall noise, routing and signal feed-through (unproperly matched)
as compared with \cite{Alberto}. The multi-peak structure observed in the time response for the
lowest voltage run ($V=5.0$ kV, $\varepsilon$=24\%) 
suggests that the noise level was abnormally high, eventually influencing
the response when the signals were very close to the threshold. This effect disappears at the
nominal voltages $V=5.6-6.0$ kV but a residual contribution to the overall resolution can certainly not
be excluded.

The chambers showed dark rates of 0.05 Hz/cm$^2$ ($V=5.6$ kV) and 0.15 Hz/cm$^2$ ($V=6.0$ kV). Few
minutes were needed after applying the HV before such values were reached, exceeding those by a factor 10-20
otherwise. Dark current was generally at the level of 1 nA or below.

\subsubsection{Dependence with particle rate}

\begin{figure}[ht!!!]
\begin{center}
\includegraphics[width = 12 cm]{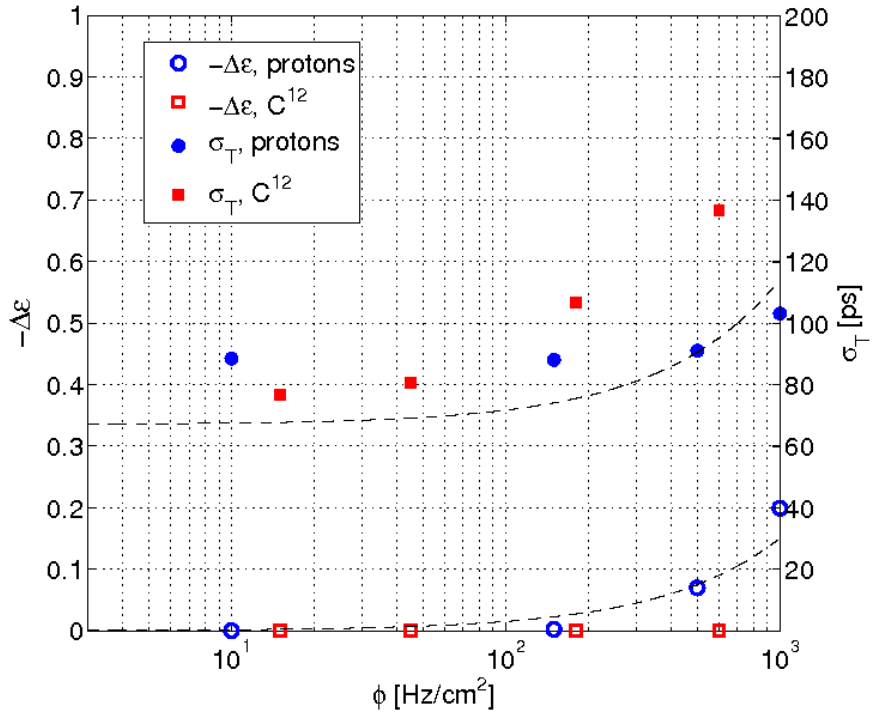}
\caption{\footnotesize Time resolution $\sigma_{_T}$ and decrease in efficiency $-\Delta{\varepsilon}$ 
in the RPCs for protons at $E_K=1.76 \pm 4\%$ GeV (circles) and fully stripped 
Carbon ions at $E_K=1.8$ GeV/A (squares)
as a function of their flux. The operating voltage was V = 5.6 kV.}
\label{rate_scan}
\end{center}
\end{figure}

The rate capability of 4-gap MtRPCs with metallic electrodes and 2 mm glass plates has been
studied under mip irradiation before \cite{HADES_Viena}, yielding a 5\% efficiency drop
($-\Delta{\varepsilon}$) at 350 Hz/cm$^2$ and slight deterioration of timing performances. Another
study in \cite{Pos_Paulo},\cite{Yo_T} consistently 
showed that a moderate $10$ $^{\circ}$C temperature increase allows to keep
the time resolution up to 1 kHz/cm$^2$ at least. We present once again the behavior of the
time resolution and efficiency as a function of the particle flux in Fig. \ref{rate_scan} for a typical
field $E=100$ kV/cm, together with the fit from previous work \cite{HADES_Viena} (dashed lines)
and added the behavior under C$^{12}$ ions. In this last case the non-uniform irradiation is
taken into account from the measured position profile (Fig. \ref{position_profile}) through
an average over $\pm 1$ cm around the peak value. For the proton case the study was performed with
a diffuse beam (section \ref{proton_beam}).

Indeed the behavior for protons as seen in Fig. \ref{rate_scan}
is similar to the one reported in \cite{HADES_Viena} but with a slightly worst resolution at low rates. Clearly,
the efficiency for C$^{12}$ is unaffected even at 600 Hz/cm$^2$ since the effective field drop is 
overcompensated by the largest initial charge (see Fig. \ref{charge_spectrum_C}). In this particular 
situation the resolution is
expected to be more sensitive to the effective field (eq. \ref{sigma_T}) through $t_{rise}(\bar{E})$ and so a
deterioration of the rate capability for C$^{12}$ by a factor $\times 5$ with respect to protons  can be
inferred from the figure (if the flux at $\sigma_{_T}=100$ ps is compared, for instance). Note that from
eq. \ref{DC} one expects (at moderate rates, when $\bar{q}_{_T}$ changes little) this factor to be 
directly the ratio of the total avalanche charges $\bar{q}_{_{T,C^{12}}}/\bar{q}_{_{T,p}}$. If the MtRPCs 
would work in the proportional regime this ratio would be $Z^2=36$.

A remarkable fact is that for MtRPCs under mip irradiation 
both $\varepsilon$ and $\sigma_{_T}$ begin to deteriorate at a similar value of
the flux while for higher initial ionizations the former
can be kept at reasonable values even at high particle fluxes and mainly $\sigma_{_T}$ is affected.

Fig. \ref{rate_scan} shows that the detection of relativistic ions up to $Z=6$ is feasible with
4-gap MtRPCs operated at typical fields ($E=100$ kV/cm), with a resolution $\sigma_{_T}\simeq{80}$ps,
and $\varepsilon\simeq100$\%. A more detailed differential study as a
function of the energy loss is presented in the next section.

\section{Dependence with particle type}
\label{Z_dependence}

\subsection{Introduction}

At fixed momentum per nucleon ($\gamma\beta$ constant), heavy charged particles lose 
energy due to electromagnetic interactions in square proportion to its electric charge
according to the Bethe-Bloch formula:
\beq
\frac{dE}{dx}=K\left(\frac{Z_m}{A_m}\right) Z^2 \frac{1}{\beta^2}\left[ \frac{1}{2} 
\ln{\frac{2m_ec^2\beta^2\gamma^2T_{max}}{I^2}} - \beta^2 \right] \label{BetheBloch}
\eeq
where $Z_m$, $A_m$ are the atomic and mass number of the medium, $m_e$ is the electron mass, $I$ the mean 
excitation energy, and $K=0.307075$ MeVg$^{-1}$cm$^2$. A prescription for calculating $Z_m/A_m$
and $I$ in mixtures according to \cite{PDG,density} will be used in the following.

Using fully stripped ions of charge $Z$ is a practical alternative to study the energy loss dependence of
a counter over a very broad dynamic range. Moreover, as long as measurements are performed in
the forward beam direction any low-Z fragment is likely to proceed from spallation reactions, 
roughly keeping its energy per nucleon ($E_K/A$) and so $\beta$. Therefore in the present experimental situation,
having a primary C$^{12}$ beam with $Z=6$, it is reasonable to assume that ions with $Z$=$1$-$6$ will have a
primary ionization in a relative proportion close to $1$, $4$, $9$, $16$, $25$, $36$. 
Additionally, for the energies
$E_K/A=1.8$ GeV used here, even in the very unlikely case that a
secondary proton (for instance) would travel forward with a kinetic
energy as low as $E_{K,p}=1/4(E_K/A)$, it would release only $20\%$ more energy 
than mips according to eq. \ref{BetheBloch}.
 The above considerations are true only for electromagnetic interactions with hadronic 
interactions expected to happen
only residually. In our setup we have no way to separate contributions from hadronic interactions but with another
reference detector placed RPC downstream, one could separate those by vetoing on it.
We will disregard the effect of hadronic interactions for the latter discussion and so, 
being $\gamma\beta\simeq2.7$ it will be useful to re-express the $Z$-dependence 
as $\Delta{E}/\Delta{E_{mips}}=Z^2$.

\subsection{Charge distribution}

\begin{figure}[ht!!!]
\begin{center}
\includegraphics[width = 6.8 cm]{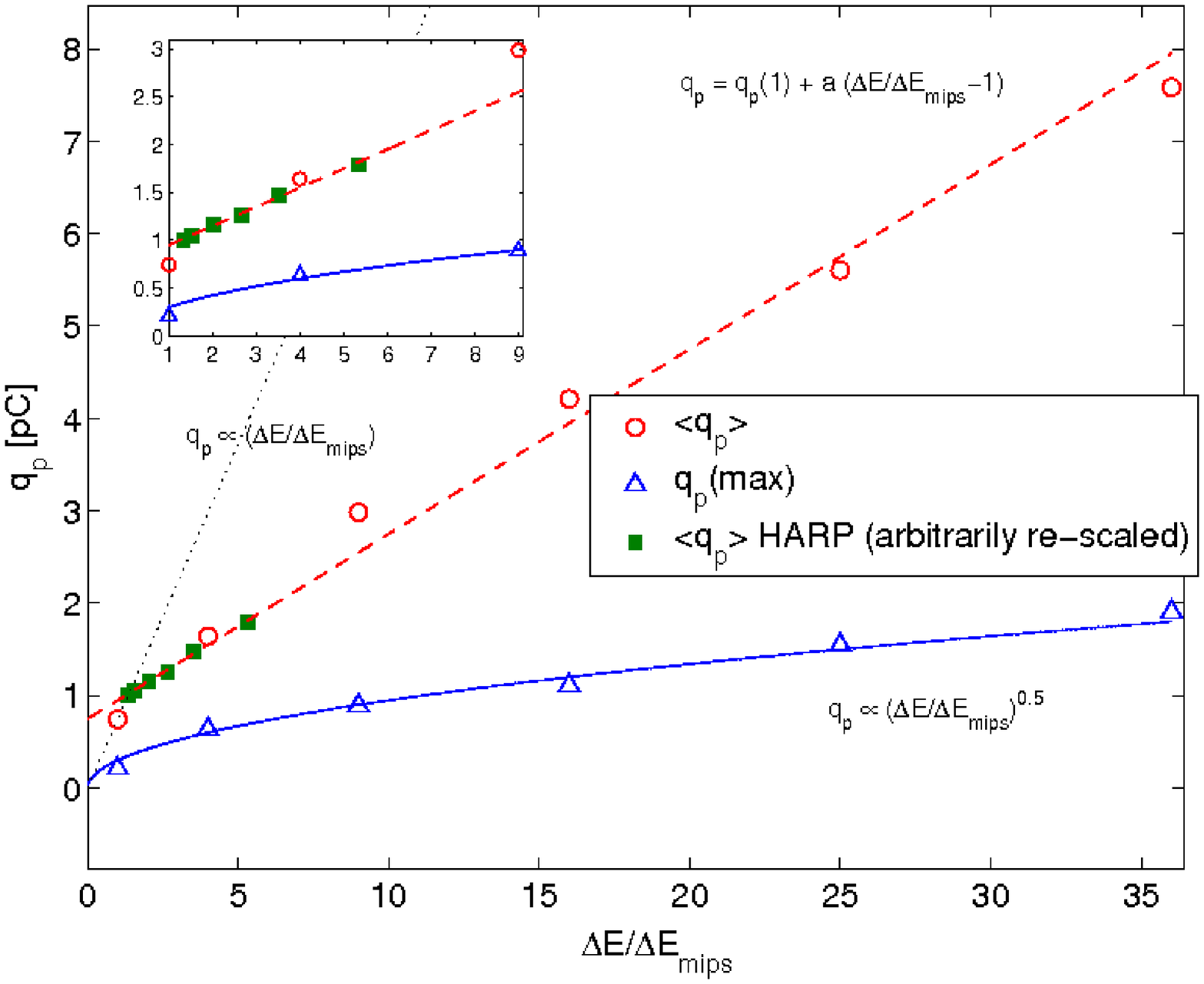}
\includegraphics[width = 6.8 cm]{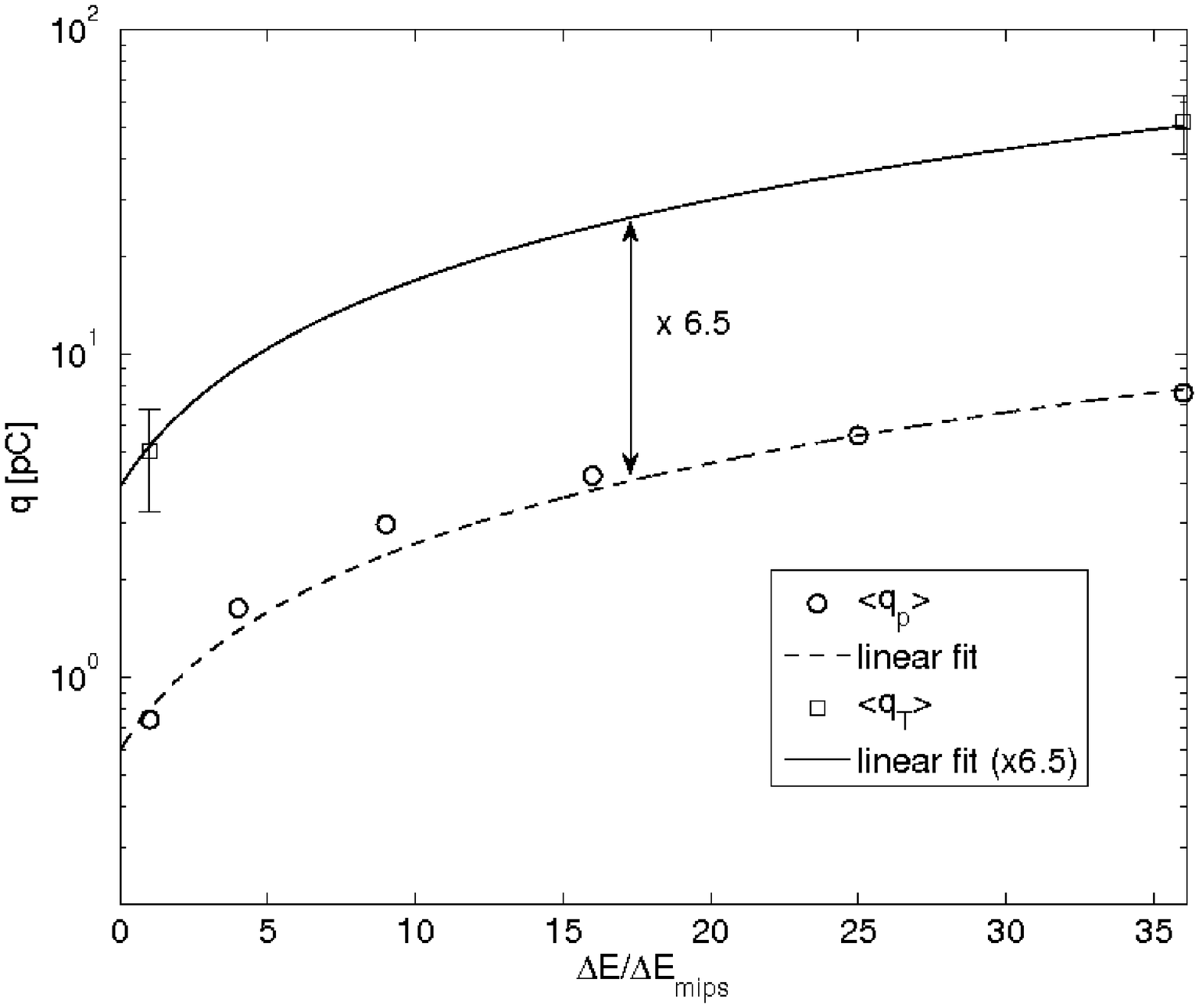}

\caption{\footnotesize Left: average prompt charge $\bar{q}_p$ (circles) and charge at maximum $q_p(\tn{max})$ 
(triangles) from the RPC 
as a function of the energy loss expressed in mips units. Data from HARP is shown by squares for comparison. 
Different trend curves are indicated. Right: average prompt $\bar{q}_p$ 
and  total $\bar{q}_{_T}$ charge (circles, squares) as a function of the energy loss expressed in mip units. 
A linear fit is shown (dashed line) together with the same curve re-scaled by a factor 6.5 
to meet the $\bar{q}_{_T}$ data.}
\label{Q_Z_scan}
\end{center}
\end{figure}

To the authors knowledge, there is only one published work on the energy loss dependence of MtRPCs \cite{HARP2},
whose conclusions yet remain disputed, specially regarding the very high shifts observed for the average
time of flight $\bar{t}$ as compared with the expected one \cite{HARP3}.
We revisited the situation starting from the dependence of the average prompt charge $\bar{q}_p$ with the energy
loss. For that we took the average charge for protons as a function of the momentum $p$ from Fig. 8
in ref. \cite{HARP2} and pad-ring 3 that corresponds to perpendicular
incidence, according to the authors. The Bethe-Bloch formula for the HARP gas mixture was 
evaluated from eq. \ref{BetheBloch} and used
for re-calculating $\Delta{E}/\Delta{x}$, re-normalizing its value to the value for mips.
Since the average prompt charge
is referred in arbitrary units in \cite{HARP2} we arbitrarily re-scaled it to provide the best agreement
with our data. Note that the geometry of the HARP cells is
very close to the HADES one, having also 4 gaps and a very similar gap spacing ($g=0.3$mm and $g=0.28$mm,
respectively). Values for HARP are shown in Fig. \ref{Q_Z_scan}-left (squares).

When increasing the energy loss the probability of streamers also increases \cite{Diego_thesis}.
The above fact makes the interpretation of the charge spectra difficult since a theoretical
description of streamers in RPCs is not yet available. In order
to simplify this task we used the fits to Landau-like functions of
Fig. \ref{charge_spectrum_C} to extract the charge at maximum $q_p(\tn{max})$
 whose energy loss dependence is expectedly dominated
by the avalanche dynamics and little dependent on the streamer mechanism. The values for $q_p(\tn{max})$
 together with the direct charge average $\bar{q}_p$ are shown in Fig. \ref{Q_Z_scan}-left. The
trends of $q_p(\tn{max})$ and $\bar{q}_p$ are 
indeed similar up to $\Delta{E}/\Delta{E_{mips}}=4$ but diverge afterwards,
pointing to the influence of streamers for  $\Delta{E}/\Delta{E_{mips}}>4$. Note the reasonable good
agreement in the trend of $\bar{q_p}$ when comparing with the HARP data. The following phenomenological
curves could be found:
\bear
&&q_p(\tn{max}) \propto \sqrt{\frac{\Delta{E}}{\Delta{E_{mips}}}} \\
&&\bar{q}_p = \bar{q}_p(1) + 0.2(\frac{\Delta{E}}{\Delta{E_{mips}}}) 
\eear
As a result of Space-Charge,
neither $q_p(\tn{max})$ nor $\bar{q}_p$ show a proportional behavior with the primary ionization.

Since the total current and particle rates were continuously monitored (Fig. \ref{time_profile})
the total average charge $\bar{q}_{_T}$ could be determined to be $\bar{q}_{_T}=5\pm{1}$ pC under proton 
irradiation and $\bar{q}_{_T}=53\pm{10}$ pC for C$^{12}$ under C$^{12}$ irradiation. The latter calculation 
requires to re-estimate the proportion of C$^{12}$ ions in
the measured rate, and is done from Fig. \ref{position_profile}, yielding 3/5 of the total. The contribution
of He and proton ions to the measured current is neglected. Note that if we assume the same scaling
with the energy loss for $\bar{q}_{_T}$ than for $\bar{q}_p$ the He and p rates (2/3 of the total) 
contribute to the measured current in less than $10\%$. From run to run variations we 
estimated in $20\%$ the error in  $\bar{q}_{_T}$ for protons and the same uncertainty was 
taken for the C$^{12}$ value.
Data for $\bar{q}_p$ and $\bar{q}_{_T}$ is shown as a function of $\Delta{E}/\Delta{E}_{mips}$ in 
Fig. \ref{Q_Z_scan}-right together with a linear fit to the $\bar{q}_p$ data. The same function 
can describe the trend of the $\bar{q}_{_T}$ points after multiplying by a factor $\times 6.5$. 
If a model of the kind of \cite{Fonte_Q} would be used for describing avalanche multiplication under 
strong Space-Charge, the ratio $\bar{q}_{_T}$/$\bar{q}_{_p}$ is indeed
expected to depend on the initial energy loss, because the higher the initial charge the earlier the
avalanche enters in the Space-Charge regime and the more the released electrons drift before arriving
to the anode, resulting in a higher electron-induced charge for the same avalanche charge. Because of that,
and because of the different behavior of $\bar{q}_p$ and $q_p(\tn{max})$ with $\Delta{E}/\Delta{E}_{mips}$,
we believe that the scaling observed in Fig. \ref{Q_Z_scan}-right is strongly affected by 
the presence of the streamers.

\subsection{Time distribution}

\begin{figure}[ht!!!]
\begin{center}
\includegraphics[width = 6.7 cm]{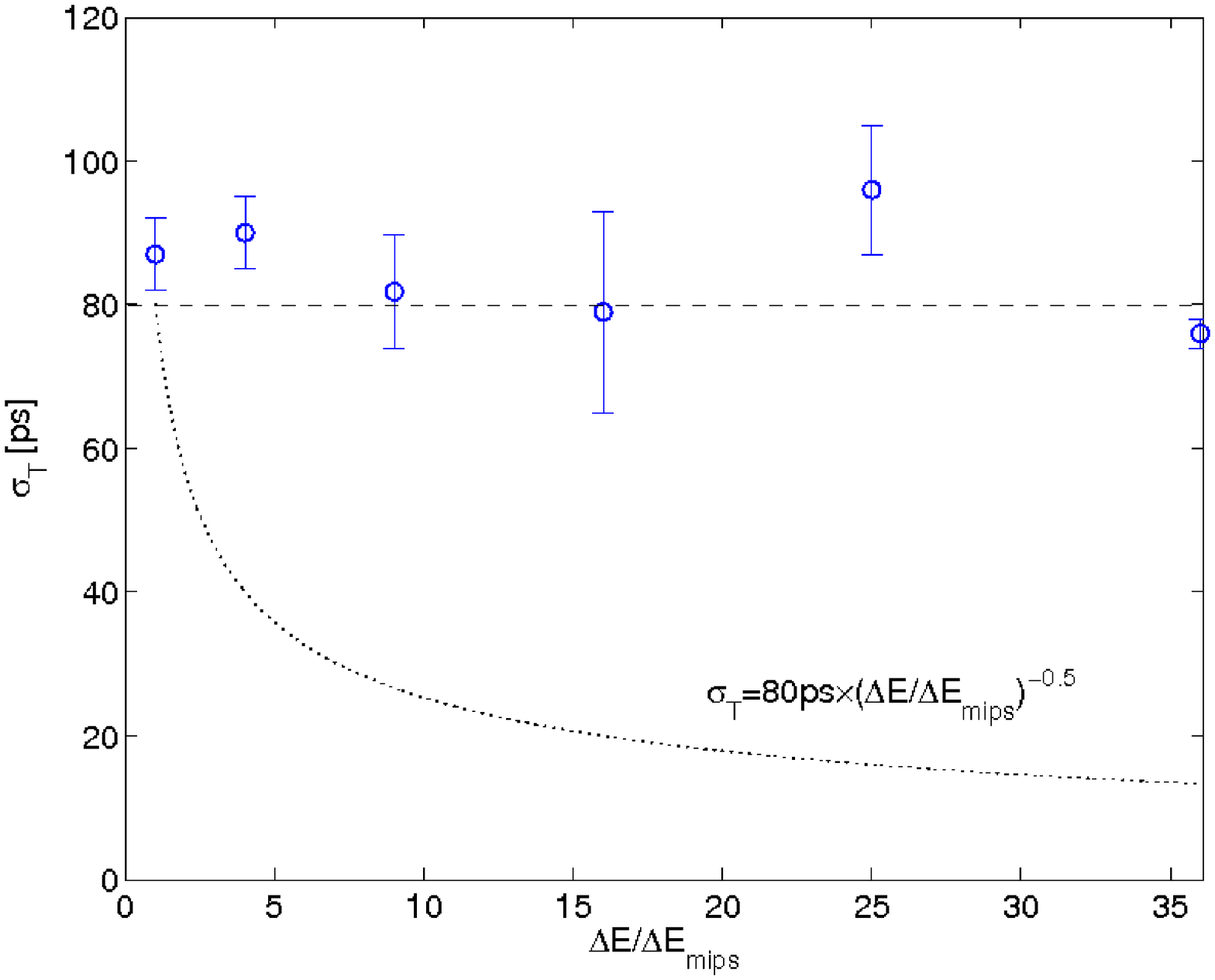}
\includegraphics[width = 6.7 cm]{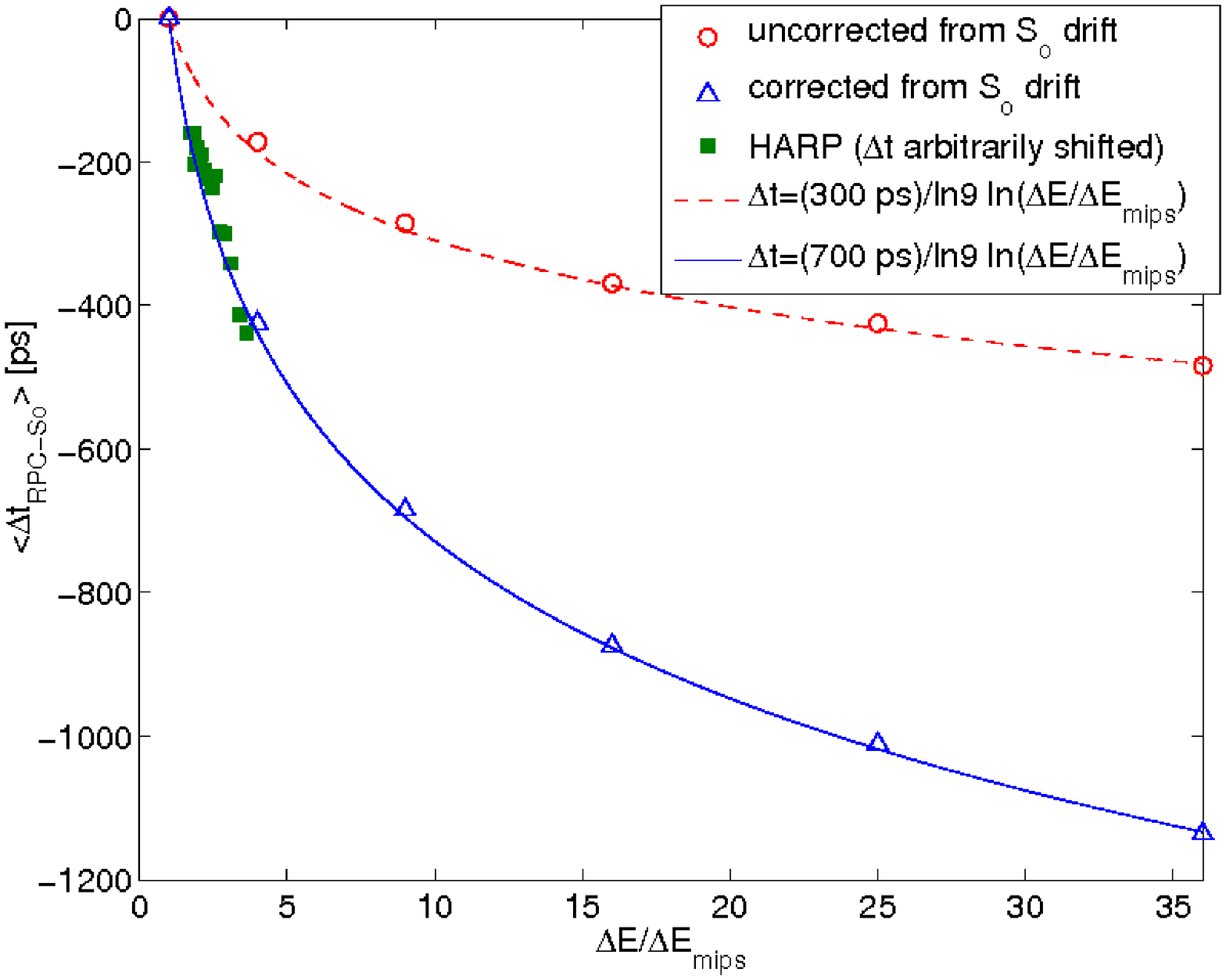}
\caption{\footnotesize Left: Time resolution $\sigma_{_T}$ as a function of the energy loss expressed 
in mips units together with the approximate $1/\sqrt{\Delta{E}}$ scaling. Right: average time of flight between RPC and scintillator $S_o$ as a function of the energy loss expressed in mips units
 together with the approximate $\ln({1/\Delta{E}})$ scaling before (circles) and after (triangles) correcting
for the scintillator drift. HARP data is shown as full squares. Two theoretical curves are fitted in each case 
with a single parameter $t_{rise}=300$ ps (dashed line) and $t_{rise}=700$ ps (continuous line).}
\label{sigma_Z_scan}
\end{center}
\end{figure}

A striking observation in \cite{HARP} was the existence of very strong drifts of the average time-of-flight 
as a function of the energy loss. The much larger range of ionizations present here allows us to discuss
this point in high detail. First, in Fig. \ref{sigma_Z_scan}-left the time resolution after slewing 
correction is shown for the different species at $V=5.6$ kV.
The analytical scaling from eq. \ref{sigma_T} is illustrated by fixing $\sigma_{_T}(1)$ to be 80 ps. 
Unfortunately, the scaling
of $\sigma_{_T}$ is not visible in data, 
pointing to the fact that a limitation to the resolution different from 
avalanche dynamics is present in the measurements.  
In connection with this observation it must be recalled that despite the solid 
derivation of eq. \ref{sigma_T} there is no systematic experimental study so far on the behavior
of $\sigma_{_T}$ with the energy loss. Proving (or disproving) the scaling of eq. \ref{sigma_T}
by systematic measurements would be extremely important in order to better understand the practical 
limitations of these counters for timing. An experimental confirmation of this scaling would allow to 
understand why, in practice, the difference between 1-gap counters and multi-gap can be as little 
as $\sigma_{_T}=55$ ps (1 gap, \cite{Paulo_1gap}) and $\sigma_{_T}=40$ ps (10 gaps, \cite{ALICE}), when 
operated in very similar conditions of gas mixture and field. No data from the HARP collaboration 
is available on this important aspect.
Since our system is limited to a resolution of $\sigma_{_T}=40$ ps for the mean-time 
(Fig. \ref{carga_uncorr}) a much more accurate FEE+TDC chain would be needed, nevertheless, in order 
to evaluate the theoretical prediction for C$^{12}$ ions.

Fig. \ref{sigma_Z_scan}-right shows the average behavior of $<\!\!\Delta{t_{RPC-S_o}}\!\!>$ (eq. \ref{deltaS0})
before corrections and referred to the case of mips (circles), together with the 
analytical scalings from eq. \ref{to} assuming $t_{rise}=300$ ps (dashed line). 
Despite the reasonable value obtained 
for $t_{rise}$ this is indeed the minimum value consistent with the observations: if a 
sizeable drift is present in the reference scintillators part of the drift originated in the RPC would 
be effectively shadowed after taking the difference. 
In the absence of constant fraction discriminators the best practical approach to 
estimate this effect in the scintillators is by looking at the $\Delta{t_{S_o-S_1}}$ distributions
when the particle releases charges
 corresponding to $Z$=$1$-$6$ in $S_o$ and keeping a narrow cut around $Z$=$1$ in $S_1$. 
The above value can be used to correct $<\!\! \Delta{t_{RPC-S_o}}\!\!>$ 
from the scintillator drifts and provide an estimate of the time drift caused by the RPC alone (triangles). 
The latest HARP data from \cite{HARP3} are also overlaid (squares) after an arbitrary shift. According
to the authors the values
have been obtained after a re-analysis based on a direct determination of the momentum $p$
by physical constraints (elastic scattering) and are expected to be free from artificial drifts 
that previously arose from an incorrect momentum reconstruction. In order to compare with present data
the energy loss was re-calculated from eq. \ref{BetheBloch}, as in previous sub-section. 
Several things can be noted: i) despite the different FEE band-widths the corrected
values for $\Delta{t_{RPC-S_o}}$ agree with latest HARP data and can be described with a logarithmic
dependence if $t_{rise}=700$ ps; ii) surprisingly, the authors in \cite{HARP3} implicitly
assumed a value for $t_{rise}=100$ ps for the discussion in the paper, presumably inspired by 
theoretical values for $(\alpha-\eta)v_e$, despite this seems to be largely inconsistent with their 
own measurements; iii) the previously reported `500 ps' effect from \cite{HARP2} can also 
not be confirmed by the present measurements; iv) a lower limit for $t_{rise}>300$ ps 
readily arises from the present data.

\begin{figure}[ht!!!]
\begin{center}
\includegraphics[width = 6.5 cm]{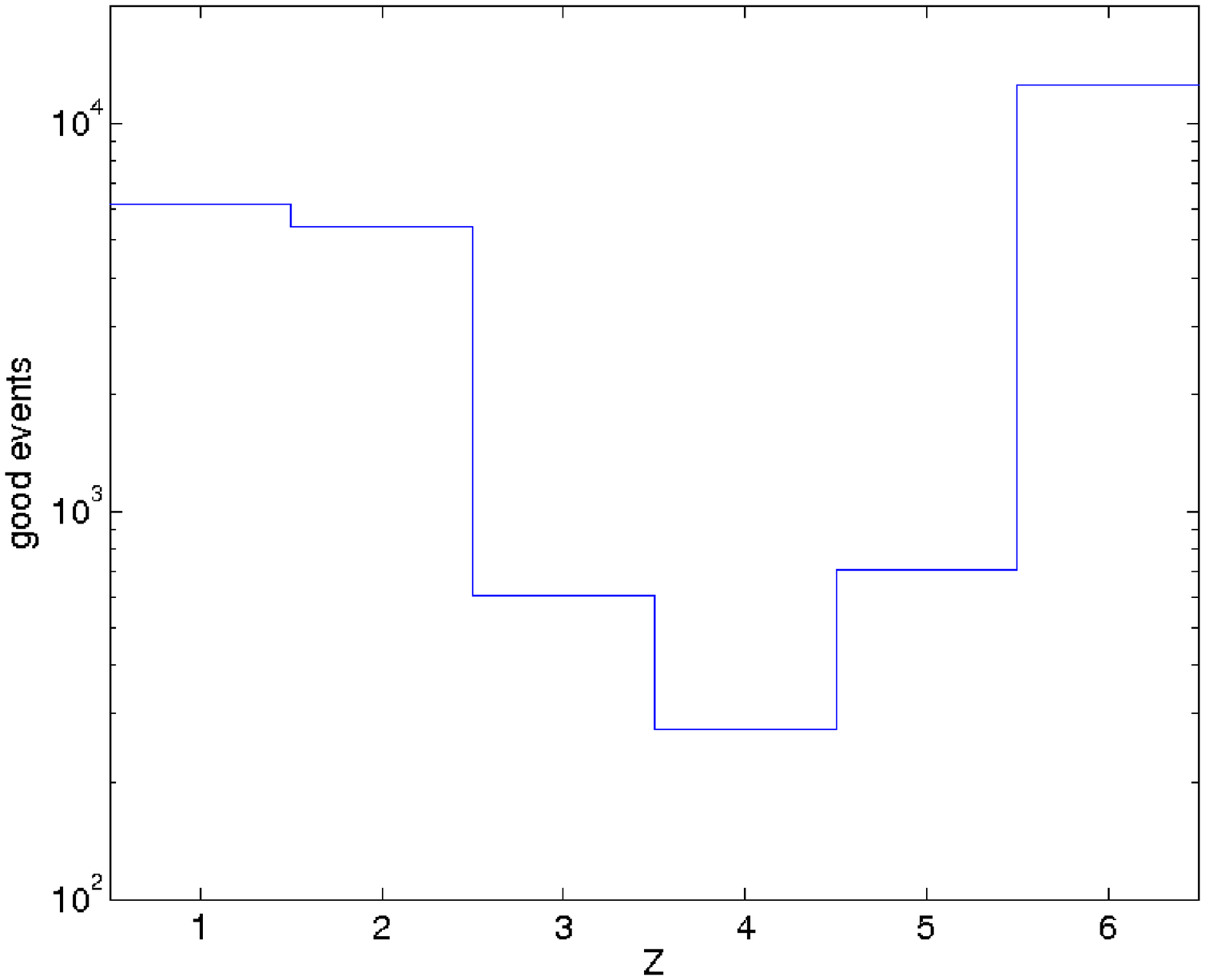}
\includegraphics[width = 6.8 cm]{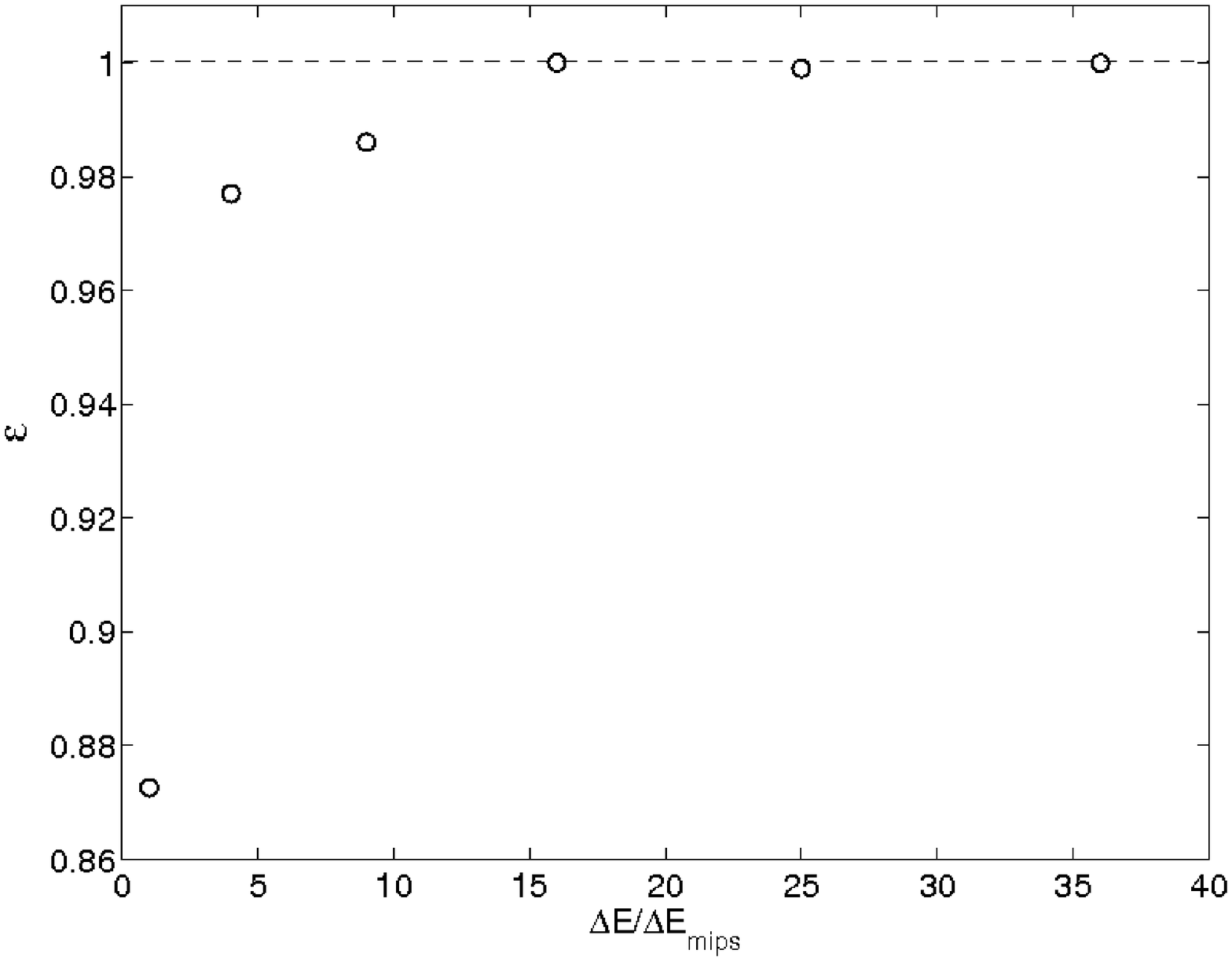}
\caption{\footnotesize Left: population of ions with charge states $1-6$ after cuts (`good events'). Right:
detection efficiency as a function of the energy loss expressed in mips units. The points have 
being obtained for $V=5.6$ kV at low particle rates ($\phi<10$ Hz/cm$^2$).}
\label{Eff_Z_scan}
\end{center}
\end{figure}

For the sake of completeness, 
fig. \ref{Eff_Z_scan}-right shows the detector efficiency as a function of the energy loss. The fact that for
$Z>1$ the efficiency is almost 100\% indicates that the trigger geometry is reasonable, being the lower
values for $Z=1$ partly attributable to the slightly low operating voltage (Fig. \ref{HV_scan}). 
Fig. \ref{Eff_Z_scan}-left presents the specie population after cuts.

\section{Discussion}
\label{Discussion}

\subsection{Element identification}

\begin{figure}[ht!!!]
\begin{center}
\includegraphics[width = 10 cm]{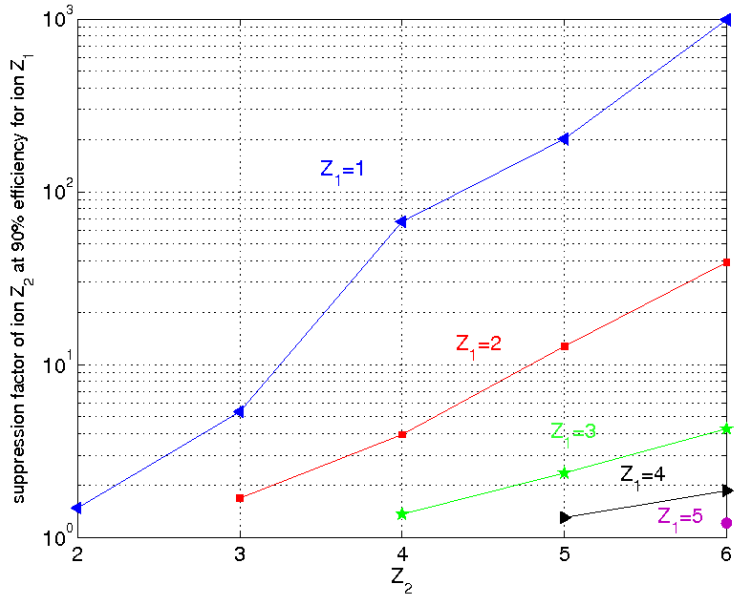}
\caption{\footnotesize Suppression factor of element in charge state $Z_2$ when imposing a cut in the 
RPC charge for a 90\% identification efficiency of element in charge state $Z_1$. The suppression factor
has been obtained after fitting the charge distributions to Landau-like functions.}
\label{Fig_purity}
\end{center}
\end{figure}

When aiming at ion identification a time of flight measurement provides the ratio 
$\gamma\beta=p/A$ while momentum ($p$) and element ($Z$) identification must be done by independent 
means \cite{Kike}. So it is interesting to see at which extent a MtRPC has resolving power in $Z$ by
using the $q_p$ information. In
order to evaluate this we define the purity $P$ against ion $Z_2$ for a 90\% identification 
efficiency of ion $Z_1$ by:
\beq
P_{90}=\left(\frac{N_{Z_1}-N_{Z_2}}{N_{Z_1}}\right)_{\varepsilon_{Z_1}=90\%}
\eeq
where the condition $\varepsilon_{Z_1}=90\%$ imposes a fixed cut in the prompt charge. 
Taking all the populations from the normalized distributions of Fig. \ref{charge_spectrum_C}
the suppression factor $\Pi_{90}$ for specie $Z_2$ when aiming at detecting $Z_1$ with 90\% efficiency
can be defined as:
\beq
\Pi_{90}=\left(\frac{N_{Z_1}}{N_{Z_2}}\right)_{\varepsilon_{Z_1}=90\%} = \frac{1}{1-P_{90}}
\eeq
$\Pi_{90}$ is shown in Fig. \ref{Fig_purity} for the different ions. 

The overall suppression is modest, but a strong suppression already exists for $Z=4$ ions 
 when aiming at proton detection ($\sim \times 65$). Contrary, the suppression of alpha particles
for proton detection as well as that of $Z=6$ ions for $Z=4$ ion detection is only slightly
higher than 1.

\subsection{Time-Charge correlation}
\label{t-q-corr}
\begin{figure}[ht!!!]
\begin{center}
\includegraphics[width = 10.5 cm]{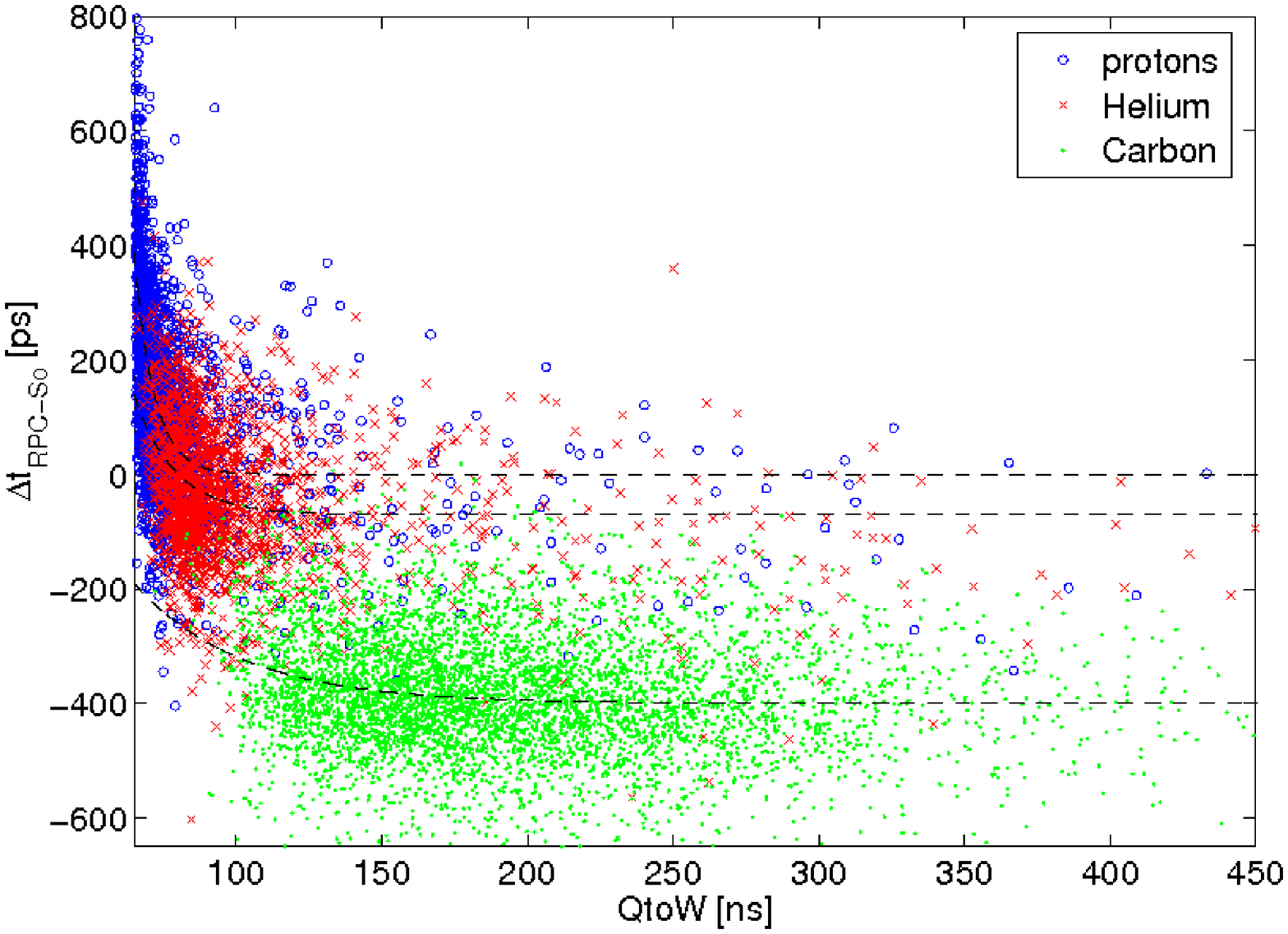}
\includegraphics[width = 10.5 cm]{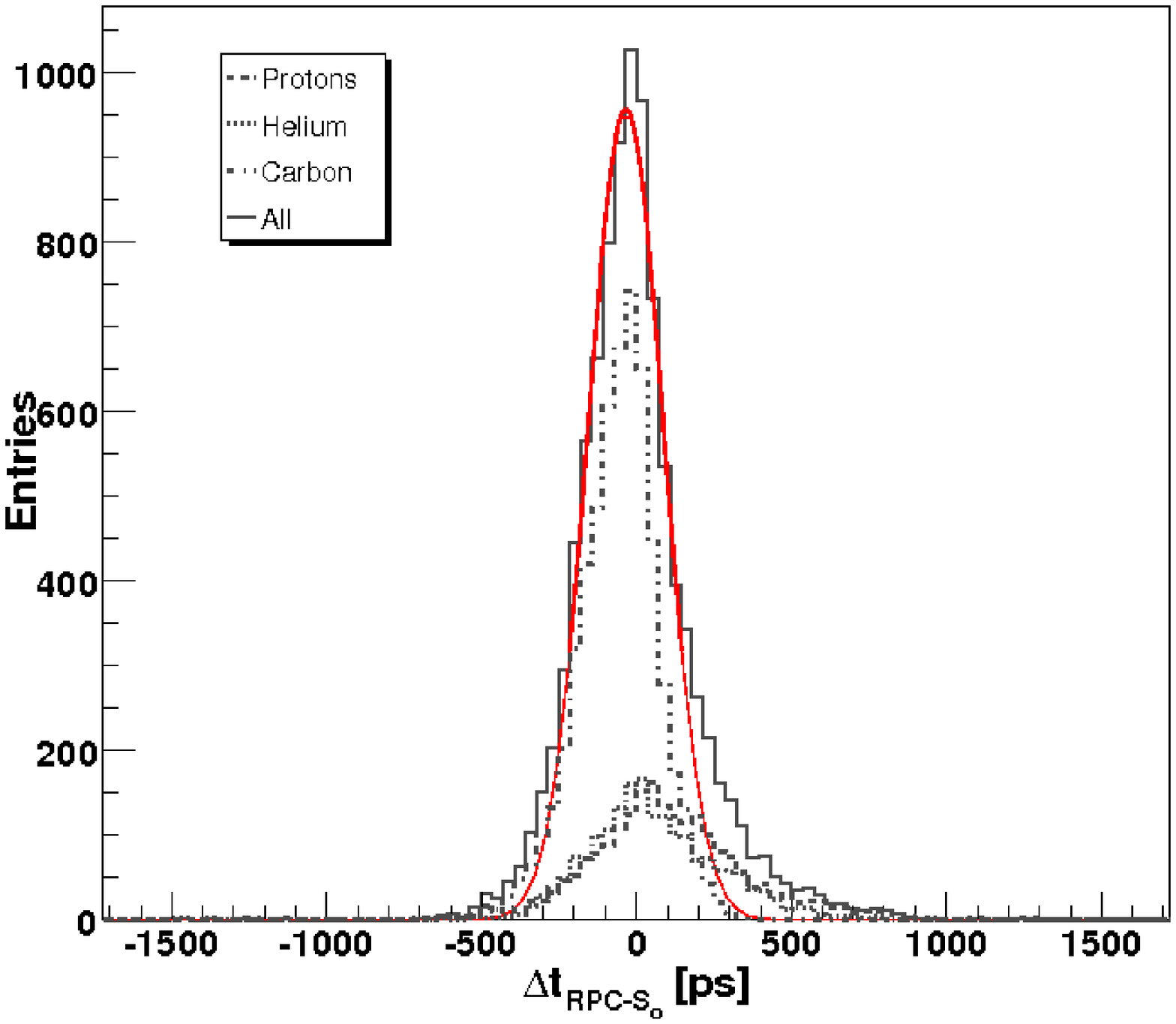}
\caption{\footnotesize Up: Scatter plot (color on-line) showing the walk 
of the time-of-flight between RPC and scintillator $S_o$ as a function of the  
signal width ($QtoW$) for protons (o), Helium (x) and Carbon ($\cdot$). Tendency curves are shown in each case.
Down: time-of-flight distribution after applying an unique slewing correction for all species 
(ignoring external identification in $Z$).
Identification is applied at the end of the procedure to evaluate the result in such conditions.
A resolution $\sigma_{_T}=115$ ps is obtained for the combined fit, with increased tails towards delayed times
as compared with the case where identification in $Z$ is provided.}
\label{Fig_corr}
\end{center}
\end{figure}

Slewing correction is the procedure after which the time of flight walk as a function of the signal
amplitude is corrected for. The nature of this correlation is not yet clear, but it seems to be dominated
by the electronic response at low charges \cite{Riegler_nice}. Indeed very little correlation, if any,
remains in the present case for $Z=6$, where the signals are largely above
threshold (Fig. \ref{Fig_corr}-up), unlike protons. Moreover, the measured time for high charges
is systematically shifted by roughly $400$ ps when comparing protons and Carbon, as 
already discussed in connection with Fig. \ref{sigma_Z_scan}. In this particular case
the external detectors provide a clean separation in $Z$ so that the
time distributions corresponding to the different ions can be identified and the drift of the average
time becomes a trivial systematic effect. Now, if $Z$ is not known (or, generally, 
if $\Delta{E}/\Delta{E}_{mips}$ is not known), this systematic dependence of the average time can not be 
fully corrected for and these variations enter effectively in the measured response function.

The worsening of the time resolution due to variations in the average time due to different primary 
ionizations is known since time \cite{Alessio2} but has only been applied to account for the Poisson
fluctuations in the initial number of clusters. Following the spirit of this work, a simple although
more experimentally-driven approach is devised here to estimate this effect in practice. Let's assume
first that the time resolution does not depend or depends little on the average ionization. This
assumption is not supported by avalanche models, but would be the case if the measured resolution
is not dominated by the detector response function but by an external source 
(FEE jitter, noise and/or TDC resolution). We think that this
case is practically relevant. Further, we assume a flat distribution of possible energy losses
over a range $\Delta{E}_{max}-\Delta{E}_{mips}$. The rms of the resulting distribution
can be estimated from the relations:
\bear
&& \bar{t}_{\Delta{E}}=\frac{1}{\Delta{E}_{max}-\Delta{E}_{mips}}\int_{\Delta{E_{mips}}}^{\Delta{E_{max}}} 
\frac{t_{rise}}{\ln{9}}\frac{\Delta{E}_{mips}}{\Delta{E}} d\Delta{E} \\
&& \sigma_{_T}^2=\frac{1}{\Delta{E}_{max}-\Delta{E}_{mips}}\int_{\Delta{E_{mips}}}^{\Delta{E_{max}}} 
\left[\left(\frac{t_{rise}}{\ln{9}}\frac{\Delta{E}_{mips}}{\Delta{E}}\right)^2 -\bar{t}_{\Delta{E}}^2 
\right]d\Delta{E} \\
\eear
that yield:
\beq
\sigma_{_T}(\Delta{E}_{max})=
\frac{t_{rise}}{\ln{9}}\sqrt{1-\frac{1}{(\frac{\Delta{E_{max}}}{\Delta{E_{mips}}}-1)^2}
\frac{\Delta{E_{max}}}{\Delta{E_{mips}}} \ln^2 \frac{\Delta{E_{max}}}{\Delta{E_{mips}}}} \label{res_deltaE}
\eeq
Eq. \ref{res_deltaE} has the curious property of tending asymptotically to $t_{rise}/\ln{9}$ for high 
energy deposition. Since the intrinsic detector resolution is expected to be of this order (eq. \ref{sigma_T})
in the limit of low ionization, one may conclude that energy spread cannot modify the detector resolution
by a large factor. In reality two things can happen: i) that the intrinsic detector resolution from 
eq. \ref{sigma_T} is reached, and then from eq. \ref{res_deltaE} 
the worsening due to the energy loss will be of the
same order than the resolution itself, being the latter partially compensated by the improvement in the intrinsic 
resolution for every value of $\Delta{E}$ (eq. \ref{sigma_T}) or ii) that it is not reached, and then
the deterioration from eq. \ref{res_deltaE} will be smaller than the resolution itself, anyway. So it seems 
that the variations on the initial ioniztion are 
called to be `second order' for the counter resolution. The above considerations
cannot account for all the possible $\Delta{E}$ distributions and must be taken with care when a highly 
Gaussian response is needed. Relevant cases where this effect should be taken into account are neutron or $\gamma$
detection due to interaction in the electrodes, where a secondary particle with its corresponding energy
distribution is released, and also ion detection. So, the usual assumption that a higher ionization renders
a better time resolution can not be considered universally true for this kind of detectors, on the basis
of the existing knowledge. A detailed discussion on this effect can be found in \cite{Lippmann_gamma}.

Another relevant situation would be the dependence of the average ionization with $Z$, assuming the
energy per nucleon to be the same for all the particle species. Let's further assume that $2N+1$
species are present in the same amount, centered around specie $\bar{Z}$. If a detector providing 
identification in $Z$ would not exist an extra jitter will appear in the form:
\bear
&& \bar{t}_{\bar{Z}} = -\frac{t_{rise}}{\ln{9}} \left( 
\sum^N_{i=1}\frac{\ln{(\bar{Z}+i)^2} + \ln{(\bar{Z}-i)^2}}{2N+1} + \frac{\ln{\bar{Z}^2}}{2N+1} \right)\\
&& \sigma_{_T}^2 = \left(\frac{t_{rise}}{\ln{9}}\right)^2 \sum^N_{i=1}\frac{\left(\ln{(\bar{Z}+i)^2} - 
\frac{\ln{9}}{t_{rise}}\bar{t}_{\bar{Z}}\right)^2 +
\left(\ln{(\bar{Z}-i)^2} - \frac{\ln{9}}{t_{rise}}\bar{t}_{\bar{Z}}\right)^2}{2N} \label{sigma_Z}
\eear
in the limit where $\bar{Z}$ and $N$ are big, but still $\bar{Z} \gg N$, eq. \ref{sigma_Z} yields:
\beq
\sigma_{_T}(N,\bar{Z})= \sqrt{\frac{8}{6}}\frac{t_{rise}}{\ln9}\frac{N}{\bar{Z}}
\eeq
so the practical influence of the uncertainty in $Z$ in the measured resolution is also little for
high $\bar{Z}$ values. However, for cases where $\bar{Z}\simeq N \simeq 1$ (the present case) the 
jitter arising from a bad identification in $Z$ will be as high as 500 ps, 
as illustrated in Fig. \ref{Fig_corr}. As a benefit, the $Z$-resolving power of the RPC itself 
will be also much higher, resulting in a time-charge correlation curve different from the usual one for 
mips (compare circles in 
Fig. \ref{Fig_corr} with all the three species together). In particular, if only Carbon
and protons would be present in our sample they can very easily identified and corrected for, and only
few events (Carbon with $QtoW<100$ ns and protons with $QtoW>100$ ns) would be wrongly identified. 
If we would have ignored for a while, 
in this particular experimental conditions, the existence of secondary
particles arising from the primary C beam, and would have used a single correlation curve the net
result would have been a deterioration of the estimated time resolution for C ions from $\sigma_{_T}=73$ ps
to $\sigma_{_T}=115$ ps, with large tails towards delayed times.

\subsection{Space-Charge}

\begin{figure}[ht!!!]
\begin{center}
\includegraphics[width = 10 cm]{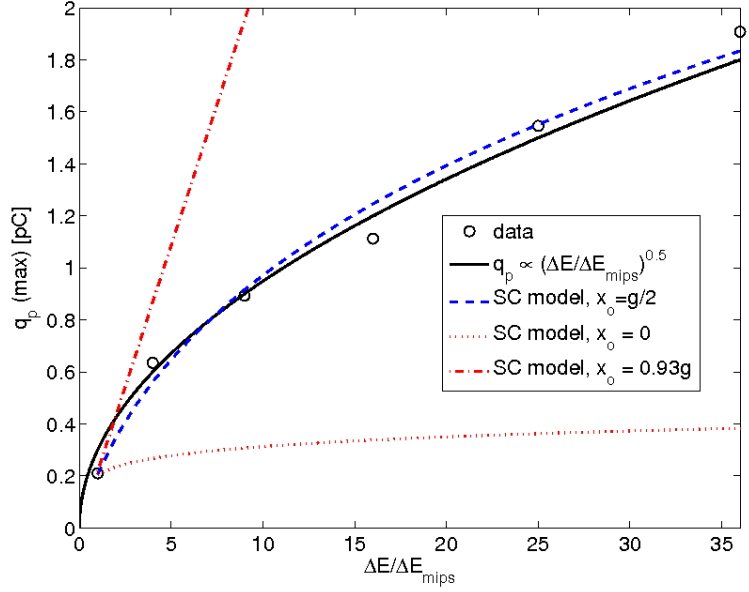}
\caption{\footnotesize Measured charge at maximum $q_p(\tn{max})$ as a function of the energy lost 
in mips units (circles). The $\sqrt{\Delta{E}}$ experimental scaling is shown together with the evaluation
of the model of Fonte \cite{Fonte_Q} under the assumptions of the text, for three different positions $x_o$
of the avalanche inside the gap (center, close to the cathode and close to the anode).}
\label{Fig_SC}
\end{center}
\end{figure}

There is a simple analytical model that accounts for Space-Charge effects, and that is 
based on the avalanche equations in the presence of a charge-dependent Townsend coefficient \cite{Fonte_Q}. 
This model is analytical
for each avalanche and depends on its position and initial ionization, but the solution to
the general problem can only be treated numerically. It is not the purpose of the present work
to quantitatively describe Space-Charge, but we did some clarifying attempt by evaluating
formulas in \cite{Fonte_Q} for three typical avalanche positions in the gap. Further, the simplifying assumption
that all the charge is released at the position of the avalanche is done. We took values 
$\alpha^* = 73$ mm$^{-1}$, $n_o$/gap$=3\Delta{E}/\Delta{E}_{mips}$, $q_{sat}=10^5$ (number of electrons 
for a 50\% drop of $\alpha^*$)
and leave the overall
normalization as a free parameter to match the data. Finally, we looked at three situations (Fig. \ref{Fig_SC}):
i) the avalanche is produced in the cathode ($x_o=0$, dotted line), ii) in 
the center of the gap ($x_o=g/2$, dashed) and iii) $20 {\mu}$m separated from the anode (dot-dashed).

Clearly, we cannot conclude from the comparison shown in 
Fig. \ref{Fig_SC} by such a simplified model, but we can illustrate how Space-Charge 
works. For avalanches produced close to the cathode there is almost no sensitivity to the initial ionization,
since its evolution is anyway doomed to a similar end once the critical value of $q_{sat}$ is reached. For 
avalanches
produced close to the anode, the resulting charge will be obviously rather small 
(an effect absorbed here in the overall normalization) but essentially proportional
to the initial one, since the avalanche charge is then much smaller than $q_{sat}$ and Space-Charge plays no role.
For the center of the gap, an intermediate situation is reached, where the avalanche growth is modified but
some sizeable correlation with the initial charge 
is still present. Remarkably, this latter situation matches well the trend of the 
experimental points.

\subsection{Streamers}

From the experimental point of view when aiming at ion detection it would be probably convenient 
to operate the detector in absence of streamers.
It is possible to perform a semi-quantitative evaluation of what is the practical implication
of this fact by using some classical arguments: 
if we denote by $\alpha^*_{c}$ the effective gain threshold at which streamers 
would arise, then following the approximate derivation of \cite{Raether} it can be seen that this value
will depend on the initial ionization as:
\beq
\alpha^*_c|_Z = \alpha^*_c|_{mips} - \frac{\ln{Z^2}}{g} 
\eeq
so, the critical value of the Townsend coefficient for an ion with charge $Z$ is naturally lower than
that of mips, and the same will happen for the field at which streamers will appear in one case and the other.
Now we can further assume a dependence of $\alpha^*(E)\simeq \frac{d\alpha}{dE}E - b$ 
(inspired by transport codes \cite{Riegler_nice}) and obtain the necessary decrease in the field 
to operate the chamber in the absence of streamers that would 
simply be, for a given ion $Z$:
\beq
E[\tn{kV/cm}]= \frac{dE}{d\alpha}\left[ \alpha^*_c|_{mips} - 1/g \ln{Z^2} + b \right]
\eeq
if we make the natural assumption $\alpha^*_c|_{mips} \simeq \alpha^*(E_o=\tn{100 kV/cm})$ then:
\beq
E_{_Z}[\tn{kV/cm}]= E_o[\tn{kV/cm}] - \frac{dE}{d\alpha}\left[\frac{\ln{Z^2}}{g}\right]
\eeq
as a function of $Z$. By directly substituting the simulated value for $\frac{dE}{d\alpha}$ \cite{Riegler_nice}
it can be expected that even for the 
most extreme case of Au detection ($Z=79$) a mere decrease of a 10\% in the field would be 
sufficient to operate the chamber in pure avalanche mode, and the anticipated decrease of performances
resulting from the lower field would be over-compensated 
if the scalings with $Z$ illustrated through eqs. \ref{sigma_T} and \ref{eq_eff} hold:
\bear
&&\sigma_{_T}(Z) \simeq \frac{K_1}{Z} \frac{t_{rise}}{\ln{9}} \\
&&\varepsilon(Z)\! >\! 1\!-\!\exp\!\left[{-n_o \left( 1-\frac{\eta}{\alpha} - 
\frac{\ln({1+\frac{(\alpha-\eta)}{E_w} n_{th}})}{\alpha g}\right) 
Z^2}\right]
\eear
since $t_{rise}$ and $\alpha$ have approximate linear dependences with the field $E$.


\section{Conclusions}
\label{Conclusion}

A time resolution $\sigma_{_T}\simeq 80$ ps at $\varepsilon\simeq$100\% has been consistently measured
for relativistic ions with charge states $Z$=$2$-$6$ for the first time by just using standard 
`off the shell' Multigap timing RPCs from the HADES wall. 

The energy loss dependence of the avalanche charge and detector-related time-of-flight systematic shifts have been 
compared with previous data and extended over a much larger range of primary ionizations, 
showing a reasonable agreement. The measured time drifts cannot
be accomodated in the existing theoretical framework unless a value for the signal rise-time
2.5 times bigger than current experimental estimates is assumed. 
The observed behavior of the prompt 
charge with the initial ionization seems to provide a stringent benchmark for the parameters of Space-Charge 
models and a simple comparison was attempted, showing a reasonable agreement.

Operation up to Carbon fluxes of 100 Hz/cm$^2$ was demonstrated 
with $\sigma_{_T}< 100$ ps and $\varepsilon\simeq$100\% under an 8-second spill (50\% duty cycle)
and $\simeq 2\times 2$ cm$^2$ irradiated area. Above 100 Hz/cm$^2$ the time resolution deteriorates
rapidly but the efficiency was kept up to 600 Hz/cm$^2$ (at least) 
due to the much higher initial ionization as compared
with mips. The behavior of the resolution as a function of the flux is similar to that of mips, when
re-scaling the rate by a factor $\times 5$. This value is very similar to the measured ratio of the total charges 
$\bar{q}_{_{T,C^{12}}}/\bar{q}_{_{T,p}}=6.5$, as expected from a simple DC modelling of the detector.
Based on this observation, 
the measured trend of $\bar{q}_{_{T}}(Z)$ can be extrapolated to high $Z$, yielding an approximate dependence
for the rate capability as $\Phi_{max}(Z)= \frac{1 \tn{kHz}}{0.2 Z^2}$, that would severely limit
the operating rate to barely 1 Hz/cm$^2$ for Au ions. Based on the present measurements, 
it is likely that the very high initial charge 
will not affect the chamber stability in such situation but it will largely reduce its rate capability. 
In a realistic application, the working voltage should be chosen taking this fact into account. 

The nature of the time-charge correlation and the practical timing limitations when the energy loss of the ionizing
particle can not be addressed by external means have been discussed. It has been shown that under reasonable
assumptions the detection of particles 
through secondary processes (in case of neutron or $\gamma$-photons, for instance)
would yield an extra time jitter of the order of the detector intrinsic resolution for mips, 
therefore  the good timing characteristics of these devices will be preserved even in such a situation. 
A practical example 
has being given based on the present data. Nevertheless, the intuition that higher ionization yields better 
results does not seem to be a trivial statement 
for these counters and every physical case should be probably addressed experimentally.

Despite the large dynamic range explored here the chamber+electronics performed stably during the 
whole experiment, underlining the superior performance of float glass Multi-gap timing RPCs in highly ionizing 
environments when high rates are not required.
Always depending on the $Z$ of the species of interest, a Multi-gap configuration does not seem to be mandatory
 for ion detection. A more practical RPC design could be probably based on 1-gap RPCs 
(single or mirrored \cite{Kike_talk}).

\acknowledgments

This work is supported by EU/FP6 contract 515876 and Xunta project PGIDT06 PXIC 296091PM (Galicia-Spain).
The authors would like to acknowledge the professional help of E. Schwab and T. Heinz.

\end{document}